\documentclass{article}

\usepackage{arxiv}

\usepackage[utf8]{inputenc}
\usepackage[T1]{fontenc}
\usepackage[hypertexnames=false, hidelinks]{hyperref}
\usepackage{url}
\usepackage{booktabs}
\usepackage{amsfonts} 
\usepackage{nicefrac}
\usepackage{microtype}

\usepackage{amsmath}
\usepackage{cite}
\usepackage[nameinlink]{cleveref}
\usepackage[]{listings}
\usepackage{empheq}
\usepackage{float}
\usepackage{graphicx}
\usepackage{siunitx}
\usepackage{xcolor}

\crefname{lstlisting}{listing}{listings}
\Crefname{lstlisting}{Listing}{Listings}

\DeclareSIUnit\pixel{px}
\DeclareSIUnit\cell{cell}

\title{Initial condition assessment for reaction-diffusion glioma growth models: A translational MRI/histology (in)validation study}

\author{
    Corentin~Martens\\
    Department of Nuclear Medicine -- H\^opital Erasme\\
    Université libre de Bruxelles\\
    Brussels, Belgium\\
    \texttt{corentin.martens@ulb.ac.be}\\
    \And
    Laetitia~Lebrun\\
    Department of Pathology -- H\^opital Erasme\\
    Université libre de Bruxelles\\
    Brussels, Belgium\\
    \And
    Christine~Decaestecker\\
    Laboratory of Image Synthesis and Analysis\\
    Universit\'e libre de Bruxelles\\
    Brussels, Belgium\\
    \And
    Thomas~Vandamme\\
    Laboratory of Image Synthesis and Analysis\\
    Universit\'e libre de Bruxelles\\
    Brussels, Belgium\\
    \And
    Yves-R\'emi~Van Eycke\\
    DIAPath -- Center for Microscopy and Molecular Imaging\\
    Universit\'e libre de Bruxelles\\
    Charleroi (Gosselies), Belgium\\ 
    \And
    Antonin~Rovai\\
    Department of Nuclear Medicine -- H\^opital Erasme\\
    Université libre de Bruxelles\\
    Brussels, Belgium\\
    \And
    Thierry~Metens\\
    Department of Radiology -- H\^opital Erasme\\
    Université libre de Bruxelles\\
    Brussels, Belgium\\
    \And
    Olivier~Debeir\\
    Laboratory of Image Synthesis and Analysis\\
    Universit\'e libre de Bruxelles\\
    Brussels, Belgium\\
    \And
    Serge~Goldman\\
    Department of Nuclear Medicine -- H\^opital Erasme\\
    Université libre de Bruxelles\\
    Brussels, Belgium\\
    \And
    Isabelle~Salmon\\
    Department of Pathology -- H\^opital Erasme\\
    Université libre de Bruxelles\\
    Brussels, Belgium\\
    \And
    Gaetan~Van Simaeys\\
    Department of Nuclear Medicine -- H\^opital Erasme\\
    Université libre de Bruxelles\\
    Brussels, Belgium\\
}

\begin{document}

\setlength{\abovedisplayskip}{20pt}
\setlength{\abovedisplayshortskip}{20pt}
\setlength{\belowdisplayskip}{20pt}
\setlength{\belowdisplayshortskip}{20pt}

\maketitle

\begin{abstract}
Diffuse gliomas are highly infiltrative tumors whose early diagnosis and follow-up usually rely on magnetic resonance imaging (MRI). However, the limited sensitivity of this technique makes it impossible to directly assess the extent of the glioma cell invasion, leading to sub-optimal treatment planing. Reaction-diffusion growth models have been proposed for decades to extrapolate glioma cell infiltration beyond margins visible on MRI and predict its spatial-temporal evolution. These models nevertheless require an initial condition, that is the tumor cell density values at every location of the brain at diagnosis time. Several works have proposed to relate the tumor cell density function to abnormality outlines visible on MRI but the underlying assumptions have never been verified so far. In this work we propose to verify these assumptions by stereotactic histological analysis of a non-operated brain with glioblastoma using a tailored 3D-printed slicer. Cell density maps are computed from histological slides using a deep learning approach. The density maps are then registered to a \textit{postmortem} MR image and related to an MR-derived geodesic distance map to the tumor core. The relation between the edema outlines visible on T2 FLAIR MRI and the distance to the core is also investigated. Our results suggest that (i) the previously suggested exponential decrease of the tumor cell density with the distance to the tumor core is not unreasonable but (ii) the edema outlines may in general not correspond to a cell density iso-contour and (iii) the commonly adopted tumor cell density value at these outlines is likely overestimated. These findings highlight the limitations of using conventional MRI to derive glioma cell density maps and point out the need of validating other methods to initialize reaction-diffusion growth models and make them usable in clinical practice.
\end{abstract}


\section{Introduction}
Gliomas are the most common primary brain tumors. Diffuse gliomas, which include its most aggressive form glioblastoma (GBM), are known to be highly infiltrative \cite{ostrom_2019}, with the presence of tumor cells reported as far as 4 cm from the gross tumor \cite{silbergeld_1997}. The early diagnosis and follow-up of gliomas usually rely on magnetic resonance imaging (MRI). However, whereas recent advents in MR technologies have given access to a significantly deeper insight into the tumor biology, none of the routinely acquired MR sequences allows to directly assess the extent of the tumor cell invasion. Instead, tumor-induced alterations of the micro-environment are seen on MR images such as peritumor vasogenic edema visible on T2/T2 FLAIR sequences and the enhancing tumor core visible on T1-weighted sequences with injection of gadolinium-based contrast agent (T1Gd). Peritumor vasogenic edema originates from an increase in the blood-brain-barrier (BBB) permeability induced by the release of vascular endothelial growth factor (VEGF) by tissues under hypoxic stress \cite{hawkins-daarud_2013, lin_2013} combined with changes in the brain hydrodynamic pressure \cite{lu_2004}. The formation of an enhancing tumor core results from a breakdown of the BBB subsequent to neo-vascularisation induced by VEGF, allowing gadolinium-based contrast agents to diffuse into brain tissues \cite{hawkins-daarud_2013}.   

The pathological and molecular examination carried out on resected or biopsied tissue samples remains the gold standard method to confirm the diagnosis and determine the histological type and grade. According to the 2016 WHO classification of the central nervous system tumors, the identification of infiltrative patterns is essential in the differential diagnosis of diffuse versus pilocytic astrocytomas, a more-circumscribed neoplasm. An overall assessment of the invasion extent would also be beneficial for surgery and radiotherapy planning. However, due to their high invasiveness and long processing times, the number and frequency of biopsy procedures are restricted, limiting the use of pathological examination as a proper tumor invasion assessment tool and highlighting the complementarity of radiological examination \cite{wesseling_2011}. In this matter, quantitatively linking glioma cell invasion patterns observed histologically to MR-visible abnormalities would be of great interest. Such information would allow to non-invasively assess glioma extent and progression, within and beyond the abnormality outlines, while providing a better interpretation of the observed abnormalities. 

Mathematical glioma growth modeling has addressed the problem of estimating glioma cell distribution within brain tissues and predicting its temporal evolution. Among the investigated models, reaction-diffusion models first introduced by Murray and colleagues in the early 1990's \cite{tracqui_1995} are probably the most widely used, with potential applications for patient follow-up and improved radiotherapy planning \cite{unkelbach_2014}. These models rely on a reaction-diffusion equation to capture the spatial-temporal evolution of a tumor cell density function:
\begin{empheq}[left=\empheqlbrace\;]{align}
    &\frac{\partial c(\boldsymbol{r}, t)}{\partial t} = \boldsymbol{\nabla} \cdot \left( \boldsymbol{D}(\boldsymbol{r}) \, \boldsymbol{\nabla} c(\boldsymbol{r}, t) \right) + \rho \, c(\boldsymbol{r}, t) \left( 1-c(\boldsymbol{r}, t) \right) & \forall \boldsymbol{r} \in \Omega, \; \forall t > 0
    \label{eq:1} \\ 
    &c(\boldsymbol{r}, 0) = c_0(\boldsymbol{r}) & \forall \boldsymbol{r} \in \Omega
    \label{eq:2} \\ 
    &\boldsymbol{D}(\boldsymbol{r}) \, \boldsymbol{\nabla} c(\boldsymbol{r}, t) \cdot \boldsymbol{n}_{\partial \Omega}(\boldsymbol{r}) = 0 & \forall \boldsymbol{r} \in \partial \Omega
    \label{eq:3}
\end{empheq}
where $c(\boldsymbol{r}, t)$ is the tumor cell density at position $\boldsymbol{r}$ and time $t$ normalized by the maximum carrying capacity $c_\text{max}$ of brain tissues ($c(\boldsymbol{r}, t) \in [0, 1], \, \forall \boldsymbol{r},t$), $\boldsymbol{D}(\boldsymbol{r})$ is the symmetric tumor cell diffusion tensor at position $\boldsymbol{r}$, $\rho$ is the tumor cell proliferation rate, $c_0(\boldsymbol{r})$ is the initial tumor cell density at position $\boldsymbol{r}$, and $\boldsymbol{n}_{\partial_{\Omega}}(\boldsymbol{r})$ is a unit normal vector pointing outwards the boundary $\partial_{\Omega}$ of the brain domain $\Omega$ at position $\boldsymbol{r} \in \partial_{\Omega}$. \Cref{eq:2} specifies the initial condition of the problem. \Cref{eq:3} provides no-flux Neumann boundary conditions reflecting the inability of tumor cells to diffuse across $\partial_{\Omega}$. 

Reaction-diffusion models as the one in \Cref{eq:1,eq:2,eq:3} are particularly attractive for clinical applications since they only have a few parameters that could be assessed from patient imaging data. For instance, based on prior observations that tumor cells rather migrate along white matter tracts, Jbabdi and colleagues proposed to derive the tumor cell diffusion tensor $\boldsymbol{D}(\boldsymbol{r})$ from diffusion tensor imaging (DTI) data \cite{jbabdi_2005}. For a more detailed overview of reaction-diffusion glioma growth modeling and its potential clinical applications, the reader is referred to \cite{tracqui_1995, clatz_2005, swanson_2008, konukoglu_2010, unkelbach_2014}.

One problem arising when attempting to solve \Cref{eq:1,eq:2,eq:3} from actual imaging data of newly diagnosed glioma patients is to estimate the initial cell density $c_0(\boldsymbol{r})$ at every location of the brain domain $\Omega$. Early works on glioma growth modeling proposed to relate MR-visible abnormalities observed at time $t$ to the tumor cell density function $c(\boldsymbol{r}, t)$. In \cite{swanson_2008}, the authors suggested to model the MR imaging process as a simple cell density threshold function, that is:
\begin{align}
    I_\text{T1Gd}(\boldsymbol{r}, t) &=
    \begin{cases}
        1 \quad &\text{if } c(\boldsymbol{r}, t) \geq c_\text{enhancing} \\
        0 \quad &\text{otherwise}
        \end{cases} 
        \label{eq:4} \\
        I_\text{T2}(\boldsymbol{r}, t) &=
    \begin{cases}
        1 \quad &\text{if } c(\boldsymbol{r}, t) \geq c_\text{edema} \\
        0 \quad &\text{otherwise}
    \end{cases}
    \label{eq:5}
\end{align}
where $I_\text{T1Gd}(\boldsymbol{r}, t)$ and $I_\text{T2}(\boldsymbol{r}, t)$ are respectively the imaging functions of the T1Gd and T2/T2 FLAIR MR sequences indicating whether the abnormality is visible at location $\boldsymbol{r}$ and time $t$ on the sequence, and $c_\text{enhancing}$ and $c_\text{edema}$ are the corresponding tumor cell density detection thresholds. Based on these assumptions, the authors suggested that the outlines of the tumor enhancing core in T1Gd images and of the vasogenic edema in T2/T2 FLAIR images would correspond to iso-contours of the tumor cell density function:
\begin{equation}
    c(\boldsymbol{r}, t) =
    \begin{cases}
    c_\text{enhancing} \quad &\text{for } \boldsymbol{r} \in \partial \Omega_\text{enhancing} \\
    c_\text{edema} \quad &\text{for } \boldsymbol{r} \in \partial \Omega_\text{edema}
    \end{cases}
    \label{eq:6}
\end{equation}
where $\partial_\text{enhancing}$ and $\partial_\text{edema}$ are respectively the enhancing core and edema outlines. The authors also suggested hypothetical values for $c_\text{enhancing}$ and $c_\text{edema}$ of 0.80 and 0.16 respectively \cite{swanson_2008}, although no rationale was provided for these values. Building upon this work, Konukoglu and colleagues proposed a fast-marching approach to construct an approximate solution of \Cref{eq:1,eq:2,eq:3} at imaging time satisfying \Cref{eq:6} \cite{konukoglu_2010}. More recently, the same group suggested that, for a spatially constant and isotropic diffusion coefficient $d_\text{white}$ and from a certain distance to the tumor core, the tumor cell density in white matter would approximately decrease exponentially with the distance $d$ to the core \cite{unkelbach_2014}:  
\begin{equation}
c(\boldsymbol{r}, t) \propto \exp\left(-\frac{d(\boldsymbol{r})}{\lambda_\text{white}}\right)
\label{eq:7}
\end{equation}
where $\lambda_\text{white}$ is the infiltration length of tumor cells in white matter given by $\sqrt{d_\text{white} / \rho}$. Provided two iso-cell density contours and a distance map to the tumor core, the value of $\lambda_\text{white}$ can theoretically be assessed using \Cref{eq:7}. A similar reasoning had been previously applied in \cite{tracqui_1995} for glioma growth modeling in computed tomography (CT) images, leading to the same expression as \Cref{eq:7} for the initial condition $c_0(\boldsymbol{r})$.

Nevertheless, these attempts to derive a tumor cell density distribution from MR images rely on the existence of cell density iso-contours in MR images and are based on unverified assumptions in \Cref{eq:4,eq:5,eq:6,eq:7}. However, as will be further discussed, the extent of vasogenic edema is known to be impacted by the administration of corticosteroids and anti-angiogenic treatments \cite{hawkins-daarud_2013}. Furthermore, as previously observed by our group in \cite{martens_2019}, the reaction-diffusion model is highly sensitive to the provided initial tumor cell density distribution $c_0(\boldsymbol{r})$. The validation of the aforementioned hypotheses is thus crucial for the model to be usable in clinical routine. In this work, we propose to verify these assumptions, as well as the value of 0.16 for $c_\text{edema}$ suggested in \cite{swanson_2008}, through a translational MRI/histology study conducted in a case of non-operated GBM. To this end, stereotactic histological analyses are performed using a 3D-printed slicer designed from \textit{antemortem} MRI data. Cell density maps are computed automatically from the scanned histological slides using a deep convolutional neural network and related to an MR-derived geodesic distance map to the tumor core. The relation between the edema outlines and the geodesic distance to the core is also investigated. Our results highlight the limitations of using routine MRI to derive glioma cell density maps and point out the need for other validated initialization methods to make reaction-diffusion growth models usable in clinical practice.

\section{Methods}
\subsection{Clinical case}
For the needs of this work, the case of a deceased 89-year-old female patient with GBM was studied retrospectively. The patient underwent an MRI examination in November 2017 in the context of a clinical frontal syndrome, which revealed a massive right-frontal expanding lesion with intense heterogeneous enhancement post-injection of gadolinium contrast agent and a necrotic core, surrounded by a large area of perilesional edema. Considering the patient’s age, a consensual decision was taken not to perform surgery and a diagnosis of GBM was made exclusively based on MRI. A corticotherapy (methylprednisolone) and a palliative chemotherapy-based treatment (temozolomide) were initiated right after diagnosis. In December 2017, after one cycle of temozolomide, the patient died of a septic shock caused by bowel perforation. An autopsy was carried out within 12 hours after death as part of which the patient's brain was collected and fixed in formalin for 24 months. Upon microscopic examination, an infiltrating glioma with high cellularity, astrocytic phenotype cells and nuclear atypia was observed along with glomeruloid vascular proliferation and areas of pseudo-palissading necrosis. All procedures performed in this study were approved by the Hospital-Faculty Ethics Committee of H\^opital Erasme (accreditation 021/406) under reference P2019/245 and are in accordance with the 1964 Helsinki declaration and its later amendments or comparable ethical standards.

\subsection{MR image acquisitions}
The T1 (TR = \SI{8}{\milli \second}, TE = \SI{2.9}{\milli \second}, TI = \SI{950}{\milli \second}, FA = \SI{8}{\degree}) and T2 FLAIR (TR = \SI{4800}{\milli \second}, TE = \SI{320}{\milli \second}, TI = \SI{1650}{\milli \second}) MR images routinely acquired at diagnosis time on a 3T Achieva scanner (Philips Healthcare, The Netherlands) -- referred to as the `\textit{in vivo}' images hereafter -- were retrospectively used in this work for registration guidance and delineation of the vasogenic edema.

Additionally, a T1 BRAVO `\textit{ex vivo}' acquisition (TR = 8.264 ms, TE = \SI{3.164}{\milli \second}, TI = \SI{450}{\milli \second}, FA = \SI{12}{\degree}) of the brain placed inside the 3D-printed slicer (see below) was performed on a 3T Signa PET/MR scanner (GE Healthcare, USA) right before slicing. It should however be noted that brain fixation has caused convergence of the white and gray matter T1 values, as reported in \cite{tovi_1992,raman_2017}. Consequently, the acquired `\textit{ex vivo}' T1 image rather has a proton-density (PD) contrast. Also note that drainage of the extracellular fluid made it impossible to delineate edema regions on \textit{postmortem} T2 FLAIR images, motivating the use of a registered \textit{antemortem} T2 FLAIR image for edema delineation hereunder.

\subsection{Slicer design and tissue sampling}
\begin{figure}[ht!]
    \centering
    \includegraphics[width=0.75\textwidth]{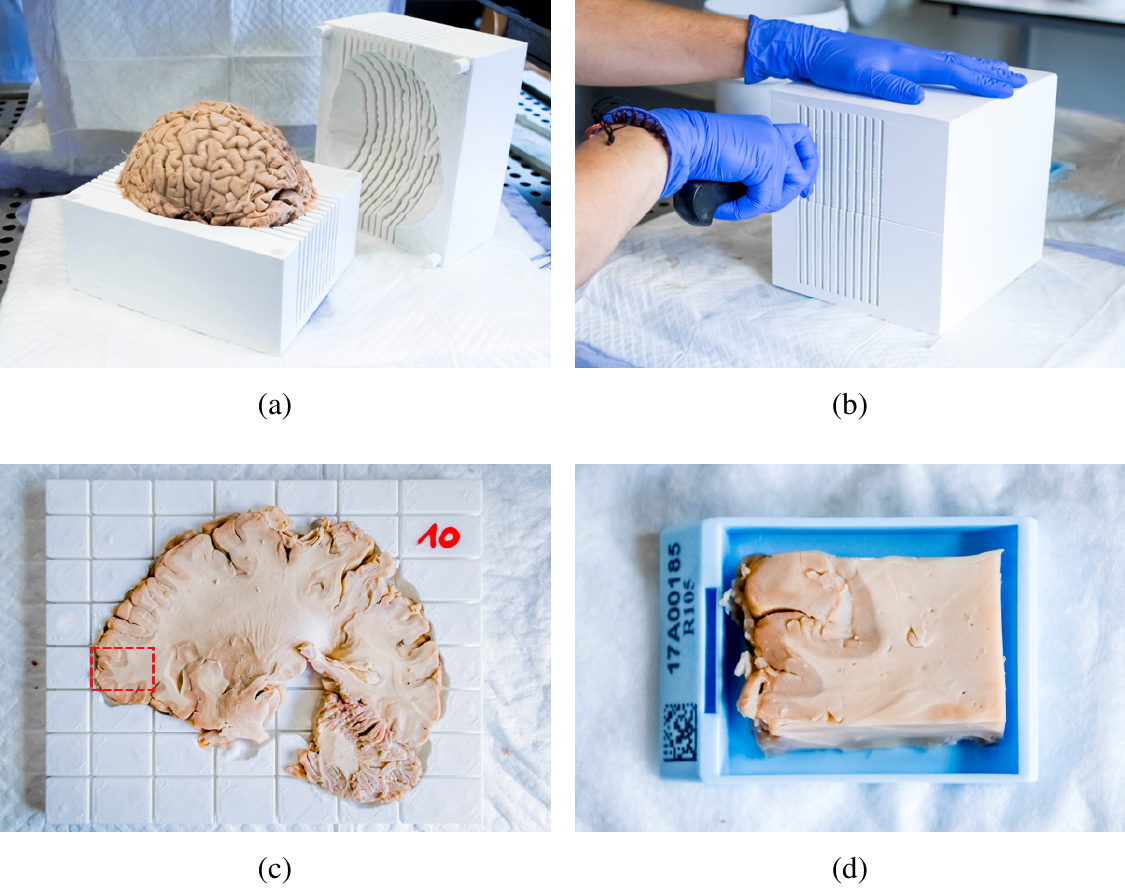}
    \caption{Brain slicing and sample collection procedure. (a) The brain is placed inside the 3D-printed slicer. (b) Sagittal slices are cut carefully. (c) Each brain slice is placed inside its cutting guide. (d) Sample blocks are cut with a scalpel along the grooves and placed into standard cassettes.}
    \label{fig:1}
\end{figure}

To relate histological observations to the abnormalities in MR images and to the MR-derived distance map to the tumor core (see below), a brain slicer was designed based on the \textit{in vivo} T2 FLAIR image and 3D-printed. Such a slicer allows the brain to be re-positioned in \textit{antemortem} imaging orientation and facilitates the cutting of sagittal brain slices. The slicer design procedure is illustrated in \Cref{fig:a1}. A similar slicer design approach was previously adopted in \cite{absinta_2014}. Ten guides were also designed to ease the collection of sample blocks from brain slices that are compatible with our histological processing chain. These consist of plates with grooves from which the brain slice volume was subtracted. The brain slicing and samples collection procedure is illustrated in \Cref{fig:1}. More details on the design steps of the slicer and the cutting guides are available in \Cref{app:a}.

\subsection{Sample processing and analysis}
28 tissue samples were selected within the tumor core, as well as within and beyond the peritumoral edematous region based on the \textit{in vivo} T2 FLAIR image. The samples were formalin‑fixed and paraffin‑embedded (FFPE, ISO 15189). \SI{5}{\micro \meter} slides were cut from each sample and stained with hematoxylin and eosin (HE). The 28 stained slides were scanned in $20\times$ mode (\SI[per-mode=symbol]{0.46}{\micro \metre \per \pixel}) on a calibrated NanoZoomer 2.0-HT digital slide scanner (Hamamatsu Photonics, Japan) for numerical processing. The stained slides were independently examined by an experienced pathologist blinded to MRI for the presence of pseudo-palisading necrosis, tumor cells (in block or infiltrating), glomeruloid vascular proliferation and edema. As will be further discussed, immunohistochemistry (IHC) staining was also investigated but did not provide satisfactory results due to over-fixation of the brain tissues.

\subsection{Cell density maps}
\begin{figure}[ht!]
    \centering
    \includegraphics[width=0.75\textwidth]{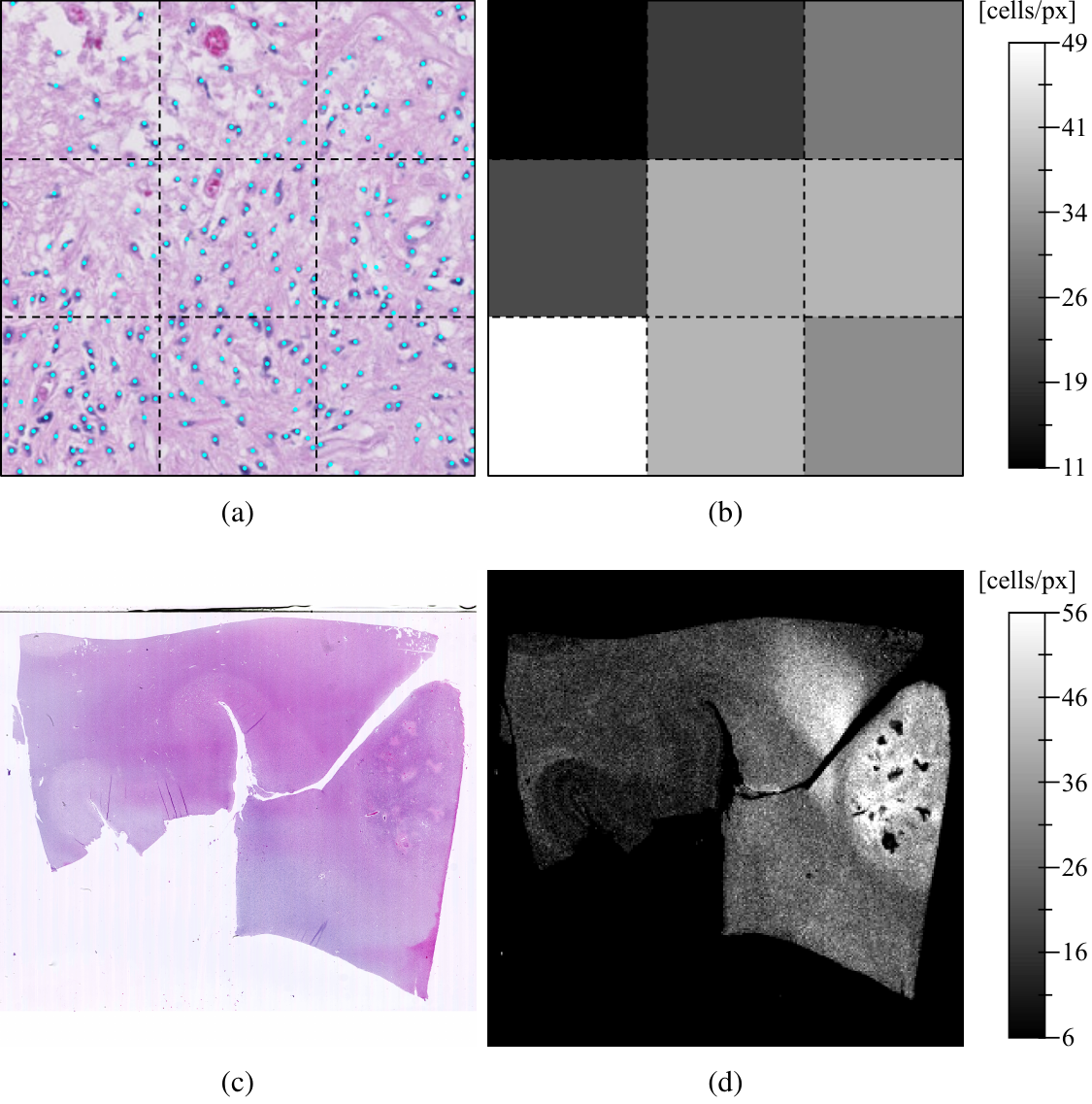}
    \caption{Cell density map computation procedure. (a) $3 \times 3$ adjacent tiles (dotted squares) with dimensions $\SI{100}{\pixel} \times \SI{100}{\pixel}$ and pixel size $\SI{1}{\micro \metre} \times \SI{1}{\micro \metre}$ extracted from the resampled slide in panel (c). Cell nuclei detected by the deep convolutional neural network are indicated with cyan dots. (b) Corresponding $3 \times 3$ pixels (dotted squares) of the cell density map with pixel size $\SI{0.1}{\milli \metre} \times \SI{0.1}{\milli \metre}$ whose value is given by the corresponding tile cell count divided by the true tissue area. (c) Whole hematoxylin and eosin stained slide (slide 13, see \Cref{tab:a1}). (d) Whole computed cell density map.}
    \label{fig:2}
\end{figure}

Cell density maps were computed from the scanned HE slides to highlight tumor cell invasion in normal brain tissues. Each scanned slide was first resampled to an isotropic pixel size of $\SI{1}{\micro \metre} \times \SI{1}{\micro \metre}$ and divided into adjacent tiles of $\SI{100}{\pixel} \times \SI{100}{\pixel}$. Cell nuclei within each tile were automatically counted using a weakly-supervised deep learning approach detailed in \Cref{app:b}. The counting result was divided by the actual tissue area within the tile, defined as the number of tissue pixels (see \Cref{app:b}) times the pixel area (\SI{e-6}{\square \milli \metre}). The computed cell density was finally stored as a 2D image with a pixel size of $\SI{0.1}{\milli \metre} \times \SI{0.1}{\milli \metre}$ where each pixel exactly corresponds to one $\SI{100}{\pixel} \times \SI{100}{\pixel}$ tile of the resampled slide. The cell density map computation procedure is illustrated in \Cref{fig:2}.

In addition, volume cell density values were extrapolated from the computed surface cell density values, since the former are the actual values of interest for the reaction-diffusion growth model. Under the assumptions that:
\begin{enumerate}
    \item Cell nuclei are approximately spherical,
    \item Cell nuclei distribution is locally isotropic,
    \item Tile dimensions are sufficiently large to contain multiple cells,
    \item Tile dimensions are sufficiently small for the cell density to be considered homogeneous and isotropic,
\end{enumerate}
and denominating $l$ the side length of the square tile, a volume cell density value $c_\text{volume}$ can be extrapolated for a cube with the same side length $l$ from the surface cell density $c_\text{surface}$ by:
\begin{equation}
    c_\text{volume} = \sqrt{c_\text{surface}}^3
    \label{eq:8}
\end{equation}

\subsection{Cell density maps to \textit{ex vivo} T1 registration}
Substantial deformations of the brain occurred between the \textit{in vivo} MR acquisitions and the histological analyses. The \textit{ex vivo} T1 image space was thus used as the reference space for the analyses and the cell density maps were registered to the \textit{ex vivo} T1 image as follows. The \textit{ex vivo} T1 image was first resampled by linear interpolation to an isotropic voxel size of \SI{0.5}{\milli \metre}. The 2D cell density maps were artificially extended to 3D by addition of a thickness of \SI{0.5}{\milli \metre} and resliced to sagittal orientation. The density maps were finally rigidly registered to the corresponding slice of the resampled \textit{ex vivo} T1 image based on user-defined landmark pairs using an in-house software in C++ based on VTK \cite{schroeder_2010} and ITK \cite{yoo_2002}.

The cell density map registration process is greatly facilitated by the use of a 3D-printed slicer since it allows to impose the brain slicing orientation. The complex histology slide to MR image registration process in 3D is thus reduced to a simple MR slice selection followed by the identification of at least 3 landmark pairs in-plane. Furthermore, the computed cell density maps have the great advantage of providing spatial tissue information at an intermediate scale between histological and radiological images, with a contrast similar to T1-weighted MR images. The cell densities of white and gray matter are indeed substantially different, as is their T1 and PD values, which eased the identification of landmarks pairs.

\subsection{Edema delineation}
\label{subsec:2_7}
To verify the assumptions in \Cref{eq:5,eq:6}, the edema region has to be delineated in the reference \textit{ex vivo} T1 image space. However, as previously mentioned, the drainage of the extracellular fluid made it impossible to discern vasogenic edema on the \textit{ex vivo} MR images. The \textit{in vivo} T2 FLAIR image was thus registered to the reference \textit{ex vivo} T1 image. To this end, the \textit{in vivo} T1 image acquired on the same day was first registered on the \textit{ex vivo} T1 image using rigid followed by B-spline transforms using the Elastix software \cite{klein_2010}. The computed transforms were then successively applied to the \textit{in vivo} T2 FLAIR image. The Elastix parameter files used for registration are available in \Cref{app:c}. The edema segmentation was finally performed semi-automatically on the registered T2 FLAIR image using a combination of thresholding and morphological operations.

\subsection{Distance map}
To verify assumption in \Cref{eq:7}, a 3D geodesic distance map to the tumor core across white matter was computed from the \textit{ex vivo} T1 image. White matter was first segmented using an in-house gradient-based anisotropic diffusion algorithm followed by manual corrections to ensure that no physically incompatible bypass exists between white matter regions. The tumor core was then segmented on the same image using a combination of thresholding and morphological operations. A distance map to the tumor core across the segmented white matter region was finally computed using an adapted implementation of the anisotropic fast marching (AFM) algorithm presented in \cite{konukoglu_2007}. Note that since no DTI images were available for the patient, the anisotropy of glioma cell diffusion mentioned hereabove could not be taken into account in this work. A unit isotropic metric tensor field was thus provided to the AFM algorithm, hence the abusive use of the term `distance map' to designate the `traveling time map' returned by the algorithm.

The relation between the edema extent and the geodesic distance map was also investigated using the Hausdorff distance and the average symmetric surface distance (ASSD), computed between the edema outlines and the contours of the binary region obtained by thresholding the distance map. The Hausdorff distance $d_\text{Hausdorff}$ and the ASSD $d_\text{ASSD}$ between two sets $A$ and $B$ are respectively given by \cite{yeghiazaryan_2018}:
\begin{gather}
d_\text{Hausdorff}(A, B) = \max\left\{ \max_{b \in B} \left\{ \min_{a \in A} d(a, b) \right\}, \max_{a \in A} \left\{ \min_{b \in B} d(a, b) \right\}\right\}
\label{eq:9} \\
\nonumber \\
d_\text{ASSD}(A, B) = \frac{1}{\lvert A \rvert + \lvert B \rvert}\left(\sum_{b \in B} \min_{a \in A} d(a, b) + \sum_{a \in A} \min_{b \in B} d(a, b)\right)
\label{eq:10}
\end{gather}
where $d(x,y)$ is the Euclidian distance between elements $x$ and $y$, and $\lvert X \rvert$ is the cardinal of set $X$.

\subsection{Cell density model}
As mentioned, the over-fixation of brain tissues prevented any IHC staining. Therefore, HE staining had to be used instead, making it difficult to distinguish infiltrating tumor cells from healthy brain cells on the scanned slides. Consequently, nuclei-based cell density maps were computed which reflect the total cell density, with no distinction between tumor and healthy cells. To address this problem, we propose to verify the following equation instead of \Cref{eq:7}:
\begin{align}
    c_\text{total}(\boldsymbol{r}) &= c_\text{tumor}(\boldsymbol{r}) + c_\text{white}
    \label{eq:11} \\
    &= c_\text{core} \exp\left(-\frac{d(\boldsymbol{r})}{\lambda_\text{white}}\right) + c_\text{white}
    \label{eq:12}
\end{align}
where $c_\text{total}$, $c_\text{tumor}$, and $c_\text{white}$ are respectively the total, tumor, and healthy cell density in white matter, and $c_\text{core}$ is the tumor cell density iso-value along the tumor core boundary. In this formulation, the cellularity of healthy white matter is supposed to be approximately constant and the invading tumor cells are assumed to be superimposed to the white matter baseline cellularity $c_\text{white}$. 

To relate the total cell density and the geodesic distance to the tumor core, the registered cell density maps were resampled to the same voxel size as the geodesic distance map ($\SI{0.5}{\milli \metre} \times \SI{0.5}{\milli \metre} \times \SI{0.5}{\milli \metre}$) and all available pairs of density/distance values among the segmented white matter voxels were extracted. $c_\text{core}$, $\lambda_\text{white}$ and $c_\text{white}$ in \Cref{eq:12} were finally least-squares fitted to the available experimental density/distance pairs using SciPy’s `optimize' module in Python \cite{virtanen_2020}.

\section{Results}
\begin{figure}[ht!]
    \centering
    \includegraphics[width=0.75\textwidth]{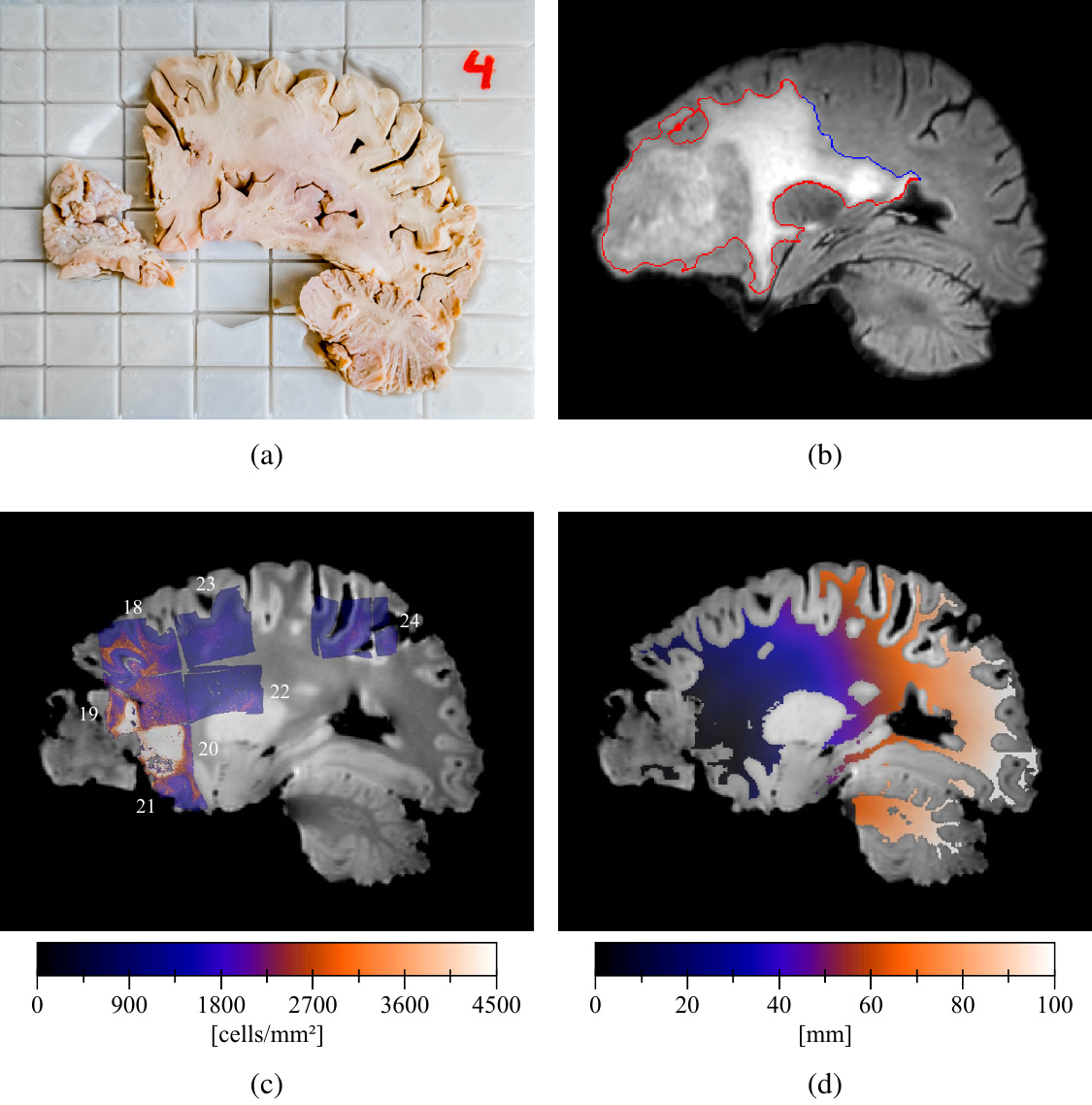}
    \caption{Cell density profile analysis. (a) Brain slice inside its 3D-printed cutting guide. (b) Corresponding slice of the registered \textit{in vivo} T2 FLAIR image with segmented edema outlines. The blue and red segments of the outline respectively correspond to free and non-free to diffuse parts of the edema boundary (see \Cref{sec:4}). (c) Corresponding slice of the \textit{ex vivo} T1 image (grayscale) and superimposed registered cell density maps (colored) with their slide number (see \Cref{tab:a1}). (d) Corresponding slice of the geodesic distance map to the tumor core across white matter.}
    \label{fig:3}
\end{figure}

\Cref{fig:3} depicts an example of brain slice in its cutting guide (\Cref{fig:3}(a)) with the corresponding registered \textit{in vivo} T2 FLAIR image slice (\Cref{fig:3}(b)), registered cell density maps (\Cref{fig:3}(c)) and geodesic distance map (\Cref{fig:3}(d)). An over-cellularity front is visible in \Cref{fig:3}(c), progressing from the frontal necrotic tumor core but rapidly decreasing to reach a normal cellularity of around \SI[per-mode=symbol]{1450}{\cell \per \square \milli \meter} beyond a geodesic distance of \SI{20}{\milli \metre} (\Cref{fig:3}(d)). The edema outlines, on the other hand, extend to over \SI{50}{\milli \metre} on the depicted image slice (see red and blue delineations in \Cref{fig:3}(b)). The distinction between the red and blue segments of the edema outlines in \Cref{fig:3}(b) will be used in the discussion.

\begin{figure}[ht!]
    \centering
    \includegraphics[width=\textwidth]{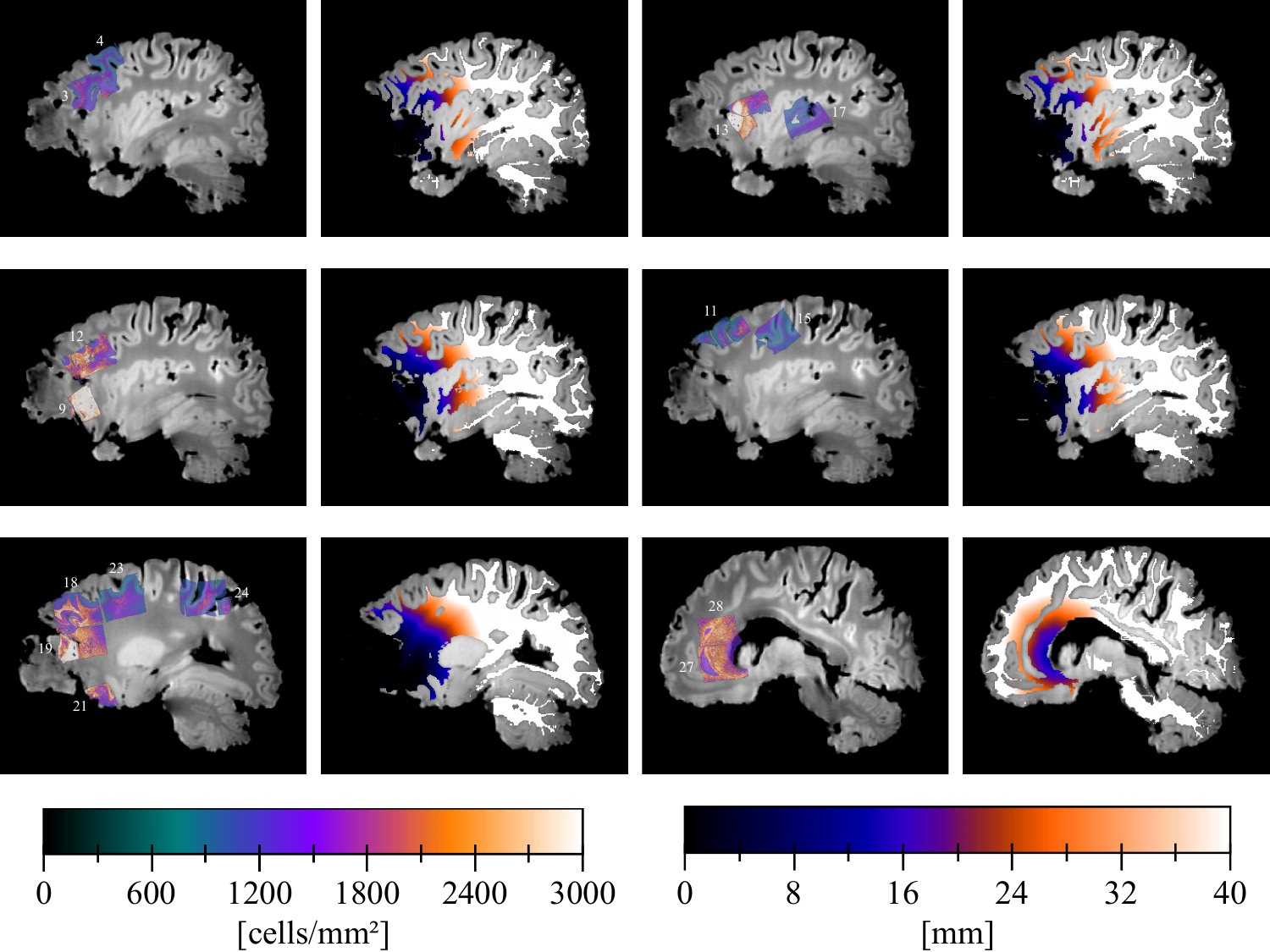}
    \caption{Example of registered cell density maps with their slide numbrt (see \Cref{tab:a1}) (1\textsuperscript{st} and 3\textsuperscript{rd} columns) and corresponding slices of the geodesic distance map to the tumor core (2\textsuperscript{nd} and 4\textsuperscript{th} columns) superimposed to the \textit{ex vivo} T1 image.}
    \label{fig:4}
\end{figure}

More examples of registered cell density maps with the corresponding geodesic distance map slice are depicted in \Cref{fig:4}. The decreasing behavior of the tumor cell density with the distance to the tumor core was observed among all these examples.

\begin{figure}[ht!]
    \centering
    \includegraphics[width=0.75\textwidth]{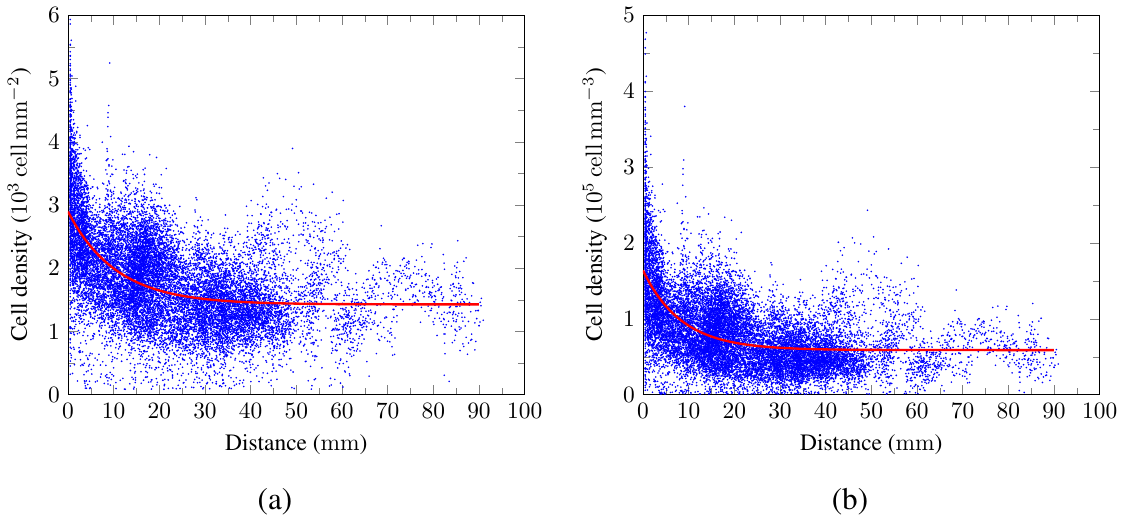}
    \caption{Scatter plot of the surface cell density (a) and the extrapolated volume cell density (b) versus distance for each available value pairs among white matter voxels (blue dots) with superimposed fitted model curves (red curves).}
    \label{fig:5}
\end{figure}

The available pairs of density/distance values among all white matter voxels are plotted in \Cref{fig:5} for the surface density data (\Cref{fig:5}(a)) and the volume density data extrapolated using \Cref{eq:8} (\Cref{fig:5}(b)). The fitted model curve given by \Cref{eq:12} is superimposed in red for each plot and the corresponding parameter values are respectively provided in \Cref{tab:1} and \Cref{tab:2}.

\begin{table}[ht!]
    \centering
    \caption{Least-squares fitted values of the cell density model parameters in \Cref{eq:12} for the surface cell density data plotted in \Cref{fig:5}(a).}
    \begin{tabular}{ccc}
        \toprule
        \textbf{$\boldsymbol{c_\text{\textbf{core}}}$ [\SI{e+3}{\cell \per \square \milli \metre}]} & \textbf{$\boldsymbol{\lambda_\text{\textbf{white}}}$ [\si{\milli \metre}]} & \textbf{$\boldsymbol{c_\text{\textbf{white}}}$ [\SI{e+3}{\cell \per \square \milli \metre}]} \\
        \midrule
        1.47 & 10.55 & 1.43 \\
        \bottomrule
    \end{tabular}
    \label{tab:1}
\end{table}

\begin{table}[ht!]
    \centering
    \caption{Least-squares fitted values of the cell density model parameters in \Cref{eq:12} for the volume cell density data extrapolated using \Cref{eq:8} and plotted in \Cref{fig:5}(b).}
    \begin{tabular}{ccc}
        \toprule
        \textbf{$\boldsymbol{c_\text{\textbf{core}}}$ [\SI{e+5}{\cell \per \cubic \milli \metre}]} & \textbf{$\boldsymbol{\lambda_\text{\textbf{white}}}$ [\si{\milli \metre}]} & \textbf{$\boldsymbol{c_\text{\textbf{white}}}$ [\SI{e+5}{\cell \per \cubic \milli \metre}]} \\
        \midrule
        1.05 & 8.46 & 0.59 \\
        \bottomrule
    \end{tabular}

    \label{tab:2}
\end{table}

\begin{figure}[ht!]
    \centering
    \includegraphics[width=0.375\textwidth]{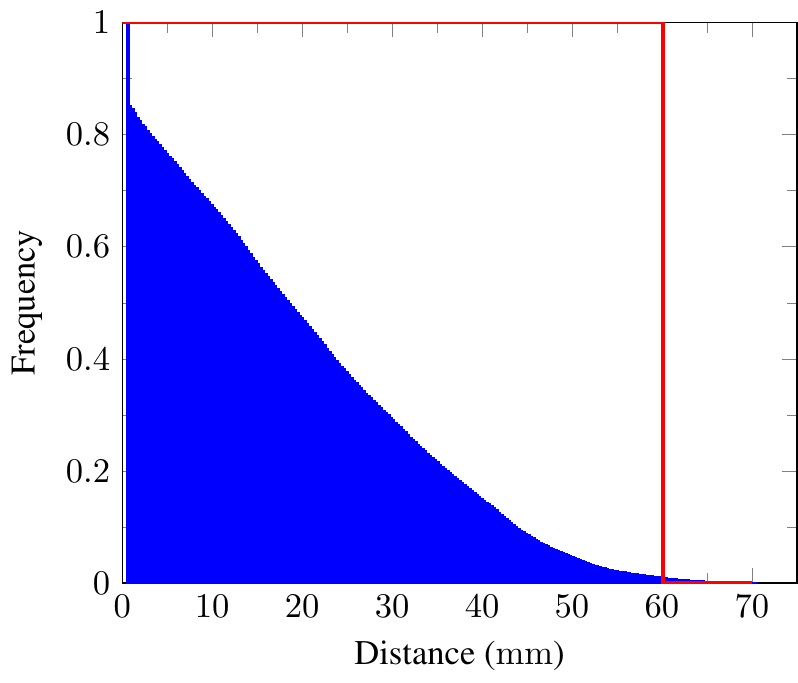}
    \caption{Inverse cumulative distribution of the geodesic distance values along the edema outlines. The expected distribution under the hypothesis of iso-distance edema outlines is plotted in red.}
    \label{fig:6}
\end{figure}

The inverse cumulative distribution of geodesic distance values along edema outlines -- i.e. the fraction of edema boundary voxels located at a geodesic distance greater or equal to a given value on the x-axis -- is depicted in \Cref{fig:6} (blue). The distribution is rather continuous, suggesting that the edema boundary does not correspond to an iso-distance contour in contrast to the expected step-like distribution superimposed in red in \Cref{fig:6}. Five percent of the edema boundary voxels are located at a distance greater or equal to \SI{49.4}{\milli \metre} and only \SI{1}{\percent} are located at a distance greater than \SI{60.1}{\milli \metre} from the tumor core.

\begin{figure}[ht!]
    \centering
    \includegraphics[width=0.75\textwidth]{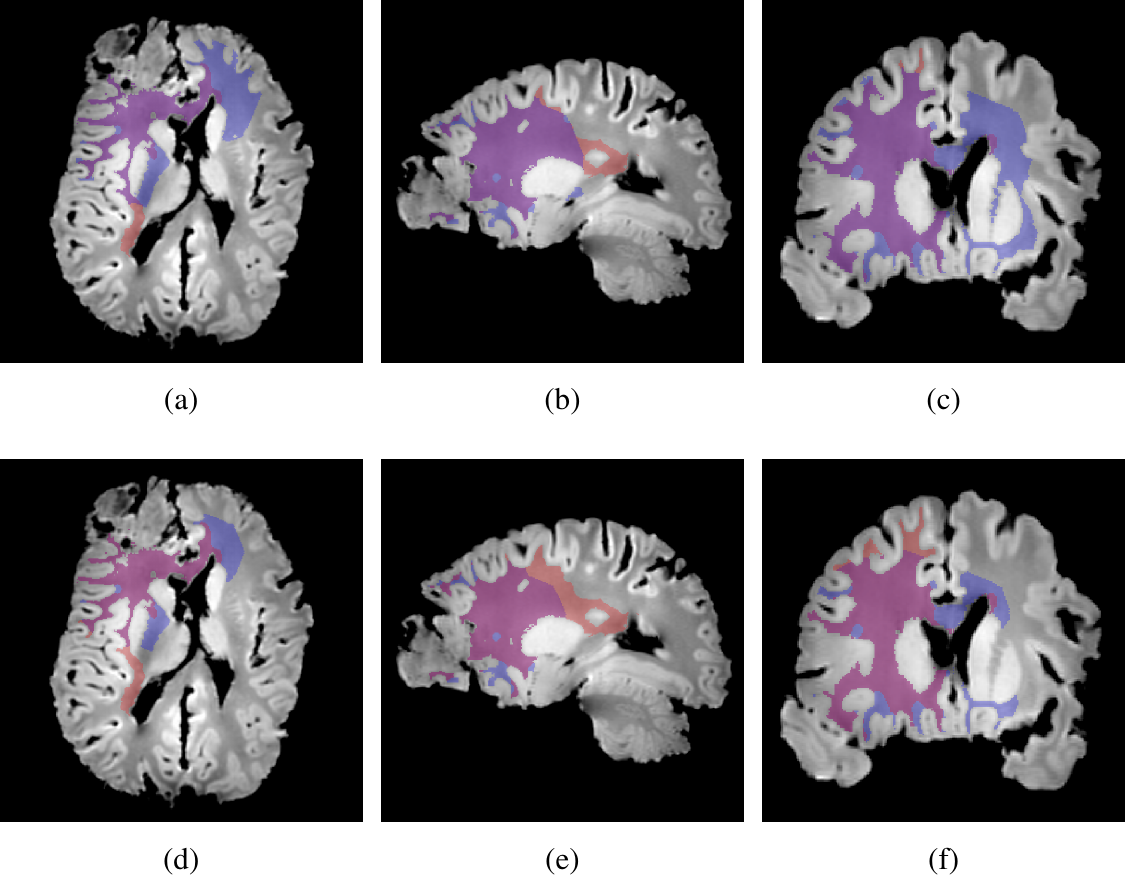}
    \caption{Edema region (red) with superimposed thresholded region of the distance map whose contour minimizes the Hausdorff distance (blue, 1\textsuperscript{st} row) and average symmetric surface distance (blue, 2\textsuperscript{nd} row) to the edema contour in axial (1\textsuperscript{st} column), sagittal (2\textsuperscript{nd} column) and coronal (3\textsuperscript{rd} column) planes.}
    \label{fig:7}
\end{figure}

The threshold geodesic distance values that provided the smallest Hausdorff distance (\SI{24.65}{\milli \metre}) and ASSD (\SI{1.97}{\milli \metre}) between the edema outlines and the contour of the corresponding thresholded region in the distance map were \SI{43.5}{\milli \metre} and \SI{35.5}{\milli \metre}, respectively. The corresponding volumes are depicted in \Cref{fig:7}.

The results of the blinded pathological examination and numerical tile processing are summarized in \Cref{tab:a1}, reporting the presence of pseudo-palisading necrosis, tumor cells (tumor block versus infiltrative cells) and glomeruloid vascular proliferation, along with the minimum, maximum and mean cell density and distance values within the corresponding slide. The furthest distance at which suspected infiltrating tumor cells were identified was around \SI{46}{\milli \meter} (slide 7), whereas edema was detected as far as \SI{61.7}{\milli \meter} on the same slide.

\section{Discussion}
\label{sec:4}
The exponentially decreasing glioma cell density profile with distance to the tumor core suggested in \cite{tracqui_1995} and \cite{unkelbach_2014} is compatible with our experimental data, as observed in \Cref{fig:5} for both surface (a) and extrapolated volume (b) cell densities. The fitted value of $\lambda_\text{white}$ (\SI{8.46}{\milli \metre}) for the volume cell density data is in the same order of magnitude as the one used in \cite{unkelbach_2014} for simulation (\SI{4.2}{\milli \metre}). The baseline volume cell density value of \SI{0.59}{\cell \per \cubic \milli \metre} in white matter is also in accordance with the literature, although large variations are to be noted between the reported values \cite{von_bartheld_2016}. However, high variance was observed in our experimental data (see blue points in \Cref{fig:5}(a) and (b)), resulting in uncertainties on the fitted parameters. The fitted values of $c_\text{core}$ were likely underestimated, since cell density values of up to \SI{6.1e+3}{\cell \per \square \milli \meter} and \SI{4.8e+5}{\cell \per \cubic \milli \meter} were respectively observed in our surface and volume cell density data (see \Cref{fig:5}(a) and (b)), and consequently the corresponding fitted values of $\lambda_\text{white}$ were probably overestimated.

In contrast, the assumption of iso-cell density edema outlines in \Cref{eq:6} was not verified by our experimental data. Indeed, considering a monotonically (exponentially) decreasing relation between the tumor cell density and the distance to the tumor core, iso-density contours should coincide with iso-distance contours. However, our results suggest that edema outlines do not coincide with an iso-distance contour (see \Cref{fig:6}) and would therefore not correspond to a cell density iso-contour either. This apparent incompatibility of assumptions in \Cref{eq:5} and \Cref{eq:6} can be explained by the thresholding behavior of the imaging function suggested in \Cref{eq:5} from which assumption in \Cref{eq:6} was deduced in \cite{swanson_2008}. In fact, thresholding a spatial function may give rise to iso-value contours only if the function is sufficiently smooth and continuous. In contrast, the tumor cell density function discretized over the voxel grid has discontinuities at interfaces between white and gray matter and along the brain domain boundary. Indeed, the difference in tumor cell diffusivity between white and gray matter \cite{unkelbach_2014,clatz_2005,konukoglu_2010} gives rise to steep tumor cell gradients at the white/gray matter interfaces, resulting in substantial discontinuities of the cell density function at the voxel precision. Along the brain domain boundary, discontinuities are even more pronounced since no tumor cell are allowed to diffuse outside the brain domain (see \Cref{eq:3}), resulting in an accumulation of cells along boundaries. Consequently, edema outlines may not correspond to cell density iso-contours even if the thresholding behavior of the edema imaging function in \Cref{eq:5} turned out to be verified. As an illustration, the edema outlines in \Cref{fig:3}(b) were split into blue and red segments, respectively corresponding to parts of the edema boundary where tumor cell diffusion is free (blue) and where diffusion is restricted due to a local decrease in tumor cell diffusivity (red). From \Cref{fig:6}, it can be reasonably assumed -- taking a margin of \SI{1}{\percent} for registration errors -- that the edema extended up to \SI{60.1}{\milli \metre} from the tumor core, which would correspond to the actual `free' edema extent. Finally, it should be noted that due to the restricted number of available cell density map voxels also belonging to the free-to-diffuse part of the edema outlines, assumption in \Cref{eq:6} could not be directly verified, which motivates the indirect distance-based reasoning herein.

For reasons mentioned hereabove, the thresholding behavior of the edema imaging function in \Cref{eq:5} may still be valid even after invalidation of \Cref{eq:6}. Nevertheless, the thresholded distance map regions whose contour respectively minimizes the Hausdorff distance and the ASSD to the edema outlines were not found to accurately coincide with the edema region, as depicted in \Cref{fig:7}. Both thresholded distance map regions (blue) in \Cref{fig:7} were indeed found to extend further in the contralateral hemisphere via the corpus callosum and less far towards the right posterior region, compared to the edema region (red). Still under the hypothesis of a monotonically decreasing relation between the tumor cell density and the distance to the tumor core, the threshold-like imaging function in \Cref{eq:6} is therefore not compatible with our results. It should however be noted that this apparent inadequacy could result from the use of an isotropic metric tensor for the geodesic map computation instead of a DTI-derived anisotropic metric tensor, which would have allowed to account for the preferential migration of glioma cells along white matter tracts \cite{jbabdi_2008}.

The iso-density value of \SI{16}{\percent} of the maximum cell carrying capacity suggested for the edema contours in \cite{swanson_2008} was not supported by our experimental data either. Indeed, assuming that \Cref{eq:6} would still be verified on the free-to-diffuse part of the edema boundary, the proposed cell density model in \Cref{eq:12} with the parameter values fitted to volume cell density data in \Cref{tab:2} suggests an over-cellularity of only \SI{8.6 e+1}{\cell \per \cubic \milli \meter} with regard to the white matter cellularity baseline $c_\text{white}$ at a distance of \SI{60.1}{\milli \metre} to the tumor core corresponding to the free edema extent. This over-cellularity corresponds to only \SI{0.08}{\percent} of $c_\text{core} \leq c_\text{max}$, which strongly invalidates the commonly accepted value of \SI{16}{\percent}. In addition, since the $c_\text{core}$ value was likely underestimated as mentioned hereabove, an even lower actual value of $c_\text{edema}$ is to be expected. These results are confirmed by blinded pathological examination, which did not reveal any noticeable invasion of the brain parenchyma -- even within the edematous region -- beyond a distance of \SI{46}{\milli \metre} to the tumor core. This overall analysis does however not exclude the possible presence of isolated infiltrating tumor cells at the edema boundary and beyond as observed in \cite{kelly_1987,silbergeld_1997,sahm_2012}.

Although MRI provides high contrast in soft tissues and is currently the standard of care for radiological examination of gliomas, abnormalities visible on conventional MR sequences are not trivially related to the tumor cell invasion extent. A striking illustration of this limitation is the administration of corticosteroid or anti-angiogenic therapies to reduce edema-related symptoms in glioma patients, which does however not stop tumor progression. Consequently, a decoupling arises between tumor progression and its visible effects on the surrounding environment, potentially leading to a misclassification of the disease as responding. The impact of such therapies on the MR-based follow-up of gliomas has been extensively studied through numerical simulations in \cite{hawkins-daarud_2013}. In this work, we invalidated two commonly made assumptions relating the outlines of visible abnormalities on MRI to the tumor cell density function: (i) Assuming a threshold-like edema imaging function (see \Cref{eq:5}), the edema contour may not correspond to an iso-contour of the cell density function as soon as the migration of tumor cells is locally restricted or prevented and (ii) at a distance corresponding to the maximum extension of the vasogenic edema, the over-cellularity was found to be negligible in our studied case, as opposed to the previously hypothesized value of \SI{16}{\percent}. These results raise the question of the applicability of both previously proposed methods to assess the glioma cell density distribution from routine MR images presented in the introduction. Indeed, whereas the method proposed in \cite{konukoglu_2010} allows to compute an accurate approximate solution of \Cref{eq:1}, it still relies on the assumption that iso-density contours can be derived from MR data. In the case of a more simple exponentially decreasing model as in \Cref{eq:7} \cite{tracqui_1995,unkelbach_2014}, iso-density contours would still be required to assess the infiltration length parameter $\lambda_\text{white}$. Since a high sensibility of the reaction-diffusion tumor growth models to their initial condition was previously reported by our group \cite{martens_2019}, deriving an initial spatial cell density distribution that is as reliable as possible is crucial for the model to be applied in clinical practice. To this end, the use of other imaging sequences or modalities such as average diffusion coefficient (ADC) maps \cite{atuegwu_2012} or amino-acid positron emission tomography should be investigated for the \textit{in vivo} assessment of the tumor cell density distribution.

This study was however prone to several limitations. First, due to the scarcity of the human body material analyzed -- a non-operated brain with GBM -- this study was based on a single case and should be further conducted on a larger diffuse glioma cohort of various grades. In addition, the use of murine glioma models for conducting such studies at a larger scale would be of interest but is restricted due to reported substantial differences in cortical \cite{hodge_2019}, glial \cite{zhang_2015}, and endothelial \cite{song_2020} cells between human and mouse brain, leading to only partial capture of human glioma features by such models \cite{de_vleeschouwer_2017}. Second, IHC staining could not be performed on the autopsied material, which has prevented the specific identification of tumor cells. Instead, HE staining was used in this work and the overall cellularity was analyzed. The assumption was made that the cellularity baseline is approximately constant across healthy white matter and that the over-cellularity observed locally is exclusively attributed to tumor cell invasion since tumor-induced recruitment of inflammatory cells is limited in brain tissues. As a consequence, the identification of isolated infiltrating tumor cells on pathological examination may have been prevented. Third, the substantial deformation of the brain between \textit{in vivo} and \textit{ex vivo} imaging -- with a volume decrease estimated to \SI{16}{\percent} \textit{ex vivo} based on MRI -- may have resulted in partial distortion of the \textit{ex vivo} MR-derived distance map compared to \textit{in vivo}, not fully compensated by deformable registration of the \textit{in vivo} edema outlines. 

We would finally like to emphasize that the automated cell density map computation and histological slide to MR image registration procedures described in this work are not limited to the problem addressed herein and could be applied to various histological stainings, imaging modalities, and organs, such as prostate. Besides, the use of tailor-made 3D-printed slicer and cutting guides makes it possible to precisely analyze whole organ slices at low cost even for centers that do not have access to whole organ slice microscopy, opening tremendous possibilities for translational microscopic/macroscopic imaging research \cite{baldi_2019}.

\section{Conclusion}
Through a translational radiological/histological analysis performed on a case of non-operated glioblastoma, we invalidated two commonly made assumptions relating the outlines of the visible abnormalities in magnetic resonance images to the tumor cell density function in the context of reaction-diffusion glioma growth modeling. We showed that, due to local restrictions of the tumor cell migration at brain tissue interfaces and along the brain boundary, the outlines of vasogenic edema in T2 FLAIR images do not generally coincide with a cell density iso-contour, as opposed to what was previously suggested. We also found that the commonly adopted tumor cell density iso-value at the edema outlines is likely overestimated since the over-cellularity at the maximum edema extent was found to be negligible in our studied case. This, however, does not exclude the possible presence of isolated tumor cells migrating beyond edema outlines, as previously reported. This work highlights the limitations of using routine magnetic resonance images to derive cell density maps for reaction-diffusion tumor growth models and points out the need of validating other methods to accurately initialize such models and make them usable for clinical applications.

\section*{Acknowledgments}
C. Martens is funded by the FRIA grant no.5120417F (F.R.S.-FNRS -- Belgian National Fund for Scientific Research). L. Lebrun is funded by the Fonds Erasme. C. Decaestecker is senior research associate with the F.R.S.-FNRS. The Department of Nuclear Medicine at H\^opital Erasme is supported by the Association Vin\c cotte Nuclear (AVN), the Fonds Erasme and the Walloon Region (Biowin). The Department of Pathology at H\^opital Erasme is supported by the Fonds Erasme and the Fonds Yvonne Bo\"el. The CMMI is supported by the European Regional Development Fund and the Walloon Region. The authors would like to thank the technical, medical, and scientific staff in charge of image acquisition and tissue sample processing for the needs of this work.

\bibliographystyle{unsrt}
\bibliography{bibliography}

\begin{thebibliography}{10}

\bibitem{ostrom_2019}
Q.~T. Ostrom, G.~Cioffi, H.~Gittleman, N.~Patil, K.~Waite, C.~Kruchko, and
  J.~S. Barnholtz-Sloan.
\newblock {CBTRUS} statistical report: Primary brain and other central nervous
  system tumors diagnosed in the {United States} in 2012–2016.
\newblock {\em Neuro-Oncol.}, 21(S5):1--100, Nov 2019.

\bibitem{silbergeld_1997}
D.~L. Silbergeld and M.~L Chicoine.
\newblock Isolation and characterization of human malignant glioma cells from
  histologically normal brain.
\newblock {\em J. Neurosurg.}, 86(3):525--31, Mar 1997.

\bibitem{hawkins-daarud_2013}
A.~Hawkins-Daarud, R.~C. Rockne, A.~R.~A. Anderson, and K.~R. Swanson.
\newblock Modeling tumor-associated edema in gliomas during anti-angiogenic
  therapy and its impact on imageable tumor.
\newblock {\em Front. Oncol.}, 3(66), Apr 2013.

\bibitem{lin_2013}
Z.-X. Lin.
\newblock Glioma-related edema: New insight into molecular mechanisms and their
  clinical implications.
\newblock {\em Chin. J. Cancer}, 32(1):49--52, Jan 2013.

\bibitem{lu_2004}
S.~Lu, D.~Ahn, G.~Johnson, M.~Law, D.~Zagzag, and R.~I. Grossman.
\newblock Diffusion-tensor {MR} imaging of intracranial neoplasia and
  associated peritumoral edema: Introduction of the tumor infiltration index.
\newblock {\em Radiology}, 232(1):221--8, Jul 2004.

\bibitem{wesseling_2011}
P.~Wesseling, J.~M. Kros, and J.~W.~M. Jeuken.
\newblock The pathological diagnosis of diffuse gliomas: Towards a smart
  synthesis of microscopic and molecular information in a multidisciplinary
  context.
\newblock {\em Diagn. Histopathol.}, 17(11):486--94, Nov 2011.

\bibitem{tracqui_1995}
P.~Tracqui, G.~C. Cruywagen, D.~E. Woodward, G.~T. Bartoo, J.~D. Murray, and
  E.~C. Alvord.
\newblock A mathematical model of glioma growth: The effect of chemotherapy on
  spatio-temporal growth.
\newblock {\em Cell Prolif.}, 28(1):17--31, Jan 1995.

\bibitem{unkelbach_2014}
J.~Unkelbach, B.~H. Menze, E.~Konukoglu, F.~Dittmann, M.~Le, N.~Ayache, and
  H.~A. Shih.
\newblock Radiotherapy planning for glioblastoma based on a tumor growth model:
  Improving target volume delineation.
\newblock {\em Phys. Med. Biol.}, 59(3):747--70, Feb 2014.

\bibitem{jbabdi_2005}
S.~Jbabdi, E.~Mandonnet, H.~Duffau, L.~Capelle, K.~Rae Swanson,
  M.~P\'el\'egrini-Issac, R.~Guillevin, and H.~Benali.
\newblock Simulation of anisotropic growth of low-grade gliomas using diffusion
  tensor imaging.
\newblock {\em Magn. Reson. Med.}, 54(3):616--24, Sep 2005.

\bibitem{clatz_2005}
O.~Clatz, M.~Sermesant, P.-Y. Bondiau, H.~Delingette, S.~K. Warfield,
  G.~Malandain, and N.~Ayache.
\newblock Realistic simulation of the {3-D} growth of brain tumors in {MR}
  images coupling diffusion with biomechanical deformation.
\newblock {\em IEEE Trans. Med. Imag.}, 24(10):1334--46, Oct 2005.

\bibitem{swanson_2008}
K.~R. Swanson, R.~C. Rostomily, and E.~C. Alvord.
\newblock A mathematical modelling tool for predicting survival of individual
  patients following resection of glioblastoma: A proof of principle.
\newblock {\em Br. J. Cancer}, 98(1):113--9, Jan 2008.

\bibitem{konukoglu_2010}
E.~Konukoglu, O.~Clatz, P.-Y. Bondiau, H.~Delingette, and N.~Ayache.
\newblock Extrapolating glioma invasion margin in brain magnetic resonance
  images: Suggesting new irradiation margins.
\newblock {\em Med. Image Anal.}, 14(2):111--25, Apr 2010.

\bibitem{martens_2019}
C.~Martens, T.~Metens, O.~Debeir, S.~Goldman, and G.~Van~Simaeys.
\newblock Initial condition assessment from patient {MRI} data for
  reaction-diffusion glioma growth models.
\newblock In {\em Proc. Intl. Soc. Mag. Reson. Med. 27}, 2019.
\newblock 2852.

\bibitem{tovi_1992}
M.~Tovi and A.~Ericsson.
\newblock Measurement of {T1} and {T2} over time in formalin-fixed human
  whole-brain specimens.
\newblock {\em Acta Radiol.}, 33(5):400--4, Sep 1992.

\bibitem{raman_2017}
M.~R. Raman, Y.~Shu, T.~G. Lesnick, C.~R. Jack, and K.~Kantarci.
\newblock Regional {T1} relaxation time constants in ex vivo human brain:
  Longitudinal effects of formalin exposure.
\newblock {\em Magn. Reson. Med.}, 77(2):774--8, Feb 2017.

\bibitem{absinta_2014}
M.~Absinta, G.~Nair, M.~Filippi, A.~Ray-Chaudhury, M.~I. Reyes-Mantilla, C.~A.
  Pardo, and D.~S. Reich.
\newblock Postmortem magnetic resonance imaging to guide the pathologic cut:
  Individualized, 3-dimensionally printed cutting boxes for fixed brains.
\newblock {\em J. Neuropathol. Exp. Neurol.}, 73(8):780--8, Aug 2014.

\bibitem{schroeder_2010}
W.~Schroeder, K.~Martin, and B.~Lorensen.
\newblock {\em The Visualization Toolkit}.
\newblock Kitware, Clifton Park, NY, USA, 4th edition, 2010.

\bibitem{yoo_2002}
T.~Yoo, M.~Ackerman, W.~Lorensen, W.~Schroeder, V.~Chalana, S.~Aylward,
  D.~Metaxas, and R.~Whitaker.
\newblock Engineering and algorithm design for an image processing {API}: A
  technical report on {ITK} – the {Insight} {Toolkit}.
\newblock {\em Stud. Health Technol. Inform.}, 85:586--92, Feb 2002.

\bibitem{klein_2010}
S.~Klein, M.~Staring, K.~Murphy, M.~A. Viergever, and J.~Pluim.
\newblock {elastix}: A toolbox for intensity-based medical image registration.
\newblock {\em IEEE Trans. Med. Imaging}, 29(1):196--205, Jan 2010.

\bibitem{konukoglu_2007}
E.~Konukoglu, M.~Sermesant, O.~Clatz, J.-M. Peyrat, H.~Delingette, and
  N.~Ayache.
\newblock A recursive anisotropic fast marching approach to reaction diffusion
  equation: Application to tumor growth modeling.
\newblock In {\em Inf. Process. Med. Imaging}, pages 687--99, Berlin,
  Heidelberg, Germany, 2007. Springer.

\bibitem{yeghiazaryan_2018}
V.~Yeghiazaryan and I.~Voiculescu.
\newblock Family of boundary overlap metrics for the evaluation of medical
  image segmentation.
\newblock {\em J. Med. Imaging}, 5(1):015006, Jan 2018.

\bibitem{virtanen_2020}
P.~Virtanen, R.~Gommers, T.~E. Oliphant, M.~Haberland, T.~Reddy, D.~Cournapeau,
  E.~Burovski, P.~Peterson, W.~Weckesser, J.~Bright, S.~J. van~der Walt,
  M.~Brett, J.~Wilson, K.~Jarrod~Millman, N.~Mayorov, A.~R.~J. Nelson,
  E.~Jones, R.~Kern, E.~Larson, C.~J. Carey, \.I. Polat, Y.~Feng, E.~W. Moore,
  J.~Vanderplas, D.~Laxalde, J.~Perktold, R.~Cimrman, I.~Henriksen, E.~A.
  Quintero, C.~R Harris, A.~M. Archibald, A.~H. Ribeiro, F.~Pedregosa, and
  P.~van Mulbregt.
\newblock {SciPy} 1.0: Fundamental algorithms for scientific computing in
  {Python}.
\newblock {\em Nat. Methods}, 17(S1):261--72, Feb 2020.

\bibitem{von_bartheld_2016}
C.~S. von Bartheld, J.~Bahney, and S.~Herculano-Houzel.
\newblock The search for true numbers of neurons and glial cells in the human
  brain: A review of 150 years of cell counting.
\newblock {\em J. Comp. Neurol.}, 524(18):3865--95, Dec 2016.

\bibitem{jbabdi_2008}
S.~Jbabdi, P.~Bellec, R.~Toro, J.~Daunizeau, M.~P\'el\'egrini-Issac, and
  H.~Benali.
\newblock Accurate anisotropic fast marching for diffusion-based geodesic
  tractography.
\newblock {\em Int. J. Biomed. Imaging}, 2008:320195, Dec 2007.

\bibitem{kelly_1987}
P.~J. Kelly, C.~Daumas-Duport, D.~B. Kispert, B.~A. Kall, B.~W. Scheithauer,
  and J.~J. Illig.
\newblock Imaging-based stereotaxic serial biopsies in untreated intracranial
  glial neoplasms.
\newblock {\em J. Neurosurg.}, 66(6):865--74, Jun 1987.

\bibitem{sahm_2012}
F.~Sahm, D.~Capper, A.~Jeibmann, A.~Habel, W.~Paulus, D.~Troost, and A.~von
  Deimling.
\newblock Addressing diffuse glioma as a systemic brain disease with
  single-cell analysis.
\newblock {\em Arch. Neurol.}, 69(4):523--6, Apr 2012.

\bibitem{atuegwu_2012}
N.~C. Atuegwu, D.~C. Colvin, M.~E. Loveless, L.~Xu, J.~C. Gore, and T.~E.
  Yankeelov.
\newblock Incorporation of diffusion-weighted magnetic resonance imaging data
  into a simple mathematical model of tumor growth.
\newblock {\em Phys. Med. Biol.}, 57(1):225--240, Jan 2012.

\bibitem{hodge_2019}
R.~D. Hodge, T.~E. Bakken, J.~A. Miller, K.~A. Smith, E.~R. Barkan, L.~T.
  Graybuck, J.~L. Close, B.~Long, N.~Johansen, O.~Penn, and Z.~Yao.
\newblock Conserved cell types with divergent features in human versus mouse
  cortex.
\newblock {\em Nature}, 573:61--8, Sep 2019.

\bibitem{zhang_2015}
Y.~Zhang, S.~A. Sloan, L.~E. Clarke, C.~Caneda, C.~A. Plaza, P.~D. Blumenthal,
  H.~Vogel, G.~K. Steinberg, M.~S.~B. Edwards, G.~Li, and J.~A.~III Duncan.
\newblock Purification and characterization of progenitor and mature human
  astrocytes reveals transcriptional and functional differences with mouse.
\newblock {\em Neuron}, 89(1):37--53, Jan 2016.

\bibitem{song_2020}
H.~W. Song, K.~L. Foreman, B.~D. Gastfriend, J.~S. Kuo, S.~P. Palecek, and
  E.~V. Shusta.
\newblock Transcriptomic comparison of human and mouse brain microvessels.
\newblock {\em Sci. Rep.}, 10(1), Dec 2020.

\bibitem{de_vleeschouwer_2017}
S.~de~Vleeschouwer, editor.
\newblock {\em Glioblastoma}, pages 131--40.
\newblock Codon Publications, Brisbane, Australia, 2017.

\bibitem{baldi_2019}
D.~Baldi, M.~Aiello, A.~Duggento, M.~Salvatore, and C.~Cavaliere.
\newblock {MR} imaging-histology correlation by tailored {3D}-printed slicer in
  oncological assessment.
\newblock {\em Contrast Media Mol. Imaging}, 2019:1071453, May 2019.

\bibitem{otsu_1979}
N.~Otsu.
\newblock A threshold selection method from gray-level histograms.
\newblock {\em IEEE Trans. Syst. Man Cybern. Syst.}, 9(1):62--6, Jan 1979.

\bibitem{ronneberger_2015}
O.~Ronneberger, P.~Fischer, and T.~Brox.
\newblock {U-Net}: Convolutional networks for biomedical image segmentation.
\newblock In {\em Med. Image Comput. Comput. Assist. Interv.}, pages 234--41,
  Cham, Switzerland, 2015. Springer.

\bibitem{abadi_2015}
M.~Abadi, A.~Agarwal, P.~Barham, E.~Brevdo, Z.~Chen, C.~Citro, G.~S. Corrado,
  A.~Davis, J.~Dean, M.~Devin, S.~Ghemawat, I.~Goodfellow, A.~Harp, G.~Irving,
  M.~Isard, R.~Jozefowicz, Y.~Jia, L.~Kaiser, M.~Kudlur, J.~Levenberg,
  D.~Man\'e, M.~Schuster, R.~Monga, S.~Moore, D.~Murray, C.~Olah, J.~Shlens,
  B.~Steiner, I.~Sutskever, K.~Talwar, P.~Tucker, V.~Vanhoucke, V.~Vasudevan,
  F.~Vi\'egas, O.~Vinyals, P.~Warden, M.~Wattenberg, M.~Wicke, Y.~Yu, and
  X.~Zheng.
\newblock {TensorFlow}: Large-scale machine learning on heterogeneous systems.
\newblock https://www.tensorflow.org/.

\bibitem{kingma_2014}
D.~Kingma and J.~Ba.
\newblock {ADAM}: A method for stochastic optimization.
\newblock {\em arXiv:1412.6980}, Dec 2014.

\bibitem{xing_2019}
F.~Xing, T.~C. Cornish, T.~Bennett, D.~Ghosh, and L.~Yang.
\newblock Pixel-to-pixel learning with weak supervision for single-stage
  nucleus recognition in {Ki67} images.
\newblock {\em IEEE Trans. Biomed. Eng.}, 66(11):3088--97, Nov 2019.

\end{thebibliography}

\appendix
\renewcommand{\theequation}{A\arabic{equation}}
\renewcommand{\thelstlisting}{A\arabic{lstlisting}}
\renewcommand{\thefigure}{A\arabic{figure}}
\renewcommand{\thetable}{A\arabic{table}}
\setcounter{equation}{0}
\setcounter{figure}{0}
\setcounter{lstnumber}{0}
\setcounter{table}{0}

\section{Slicer design}
\label{app:a}
\begin{figure}[ht!]
    \centering
    \includegraphics[width=0.75\textwidth]{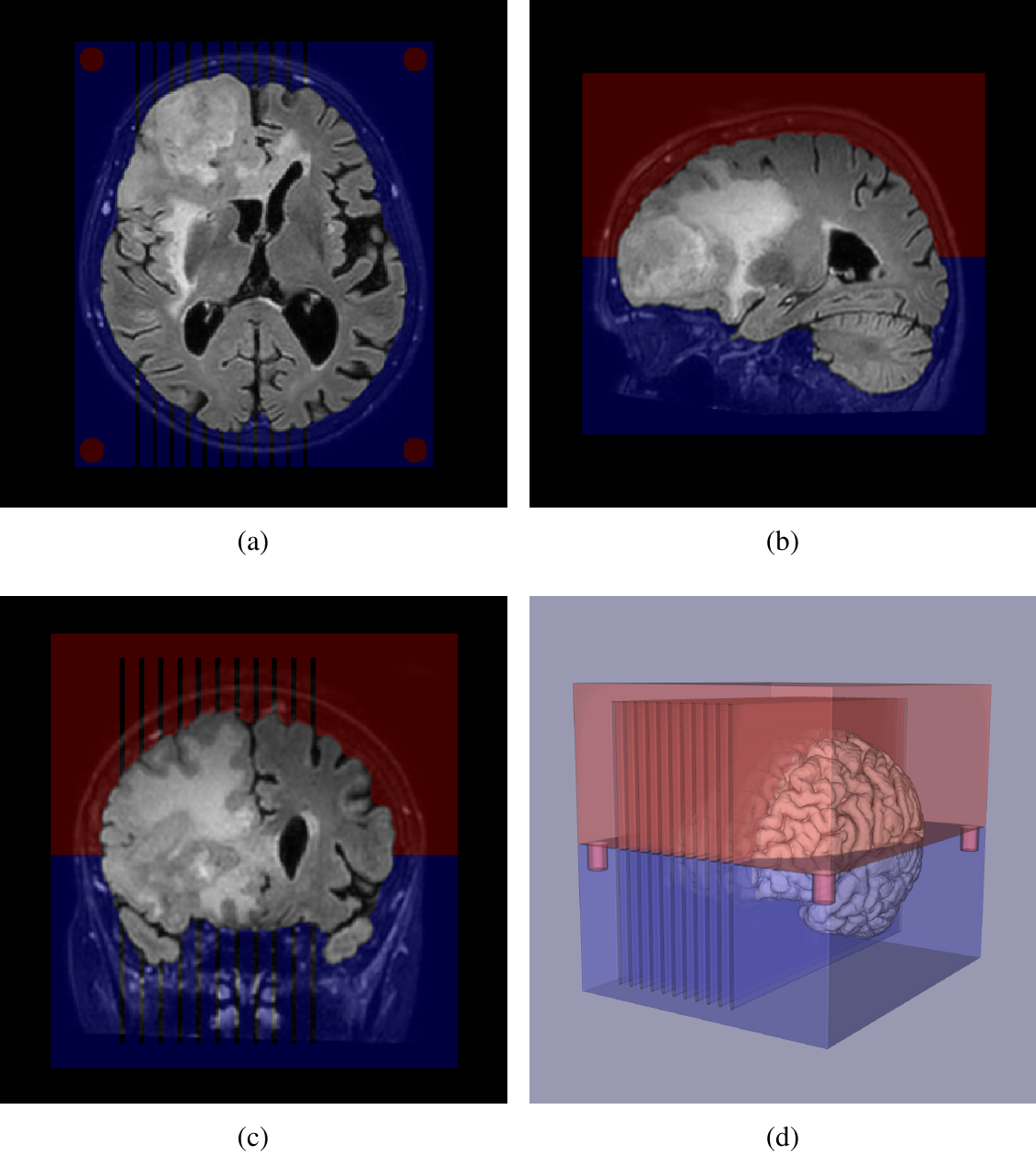}
    \caption{Brain slicer design superimposed to the \textit{in vivo} T2 FLAIR image used as template in axial (a), sagittal (b), coronal (c) and 3D (d) views.}
    \label{fig:a1}
\end{figure}

The \textit{in vivo} T2 FLAIR image was first resampled to an isotropic voxel size of $\SI{0.5}{\milli \metre} \times \SI{0.5}{\milli \metre} \times \SI{0.5}{\milli \metre}$. A brain mask was generated using the Otsu thresholding \cite{otsu_1979}, followed by a morphological opening of radius 1, a largest component extraction and a morphological closing of radius 15. The brain mask bounding box was then drawn on a binary image with the same spacing and was evenly extended on both sides in each spatial dimension, with final dimensions of $\SI{160.5}{\milli \metre} \times \SI{190.5}{\milli \metre} \times \SI{172}{\milli \metre}$. The brain mask volume was subtracted from the box volume. Ten slits of $\SI{12}{\milli \metre} \times \SI{190.5}{\milli \metre} \times \SI{152}{\milli \metre}$ spaced by \SI{5.5}{\milli \metre} were carved into the box along the sagittal plane for brain slicing and the box was split in half along the axial plane at the level of the largest brain mask outline for brain insertion. Four cylindric fixations of radius \SI{5.5}{\milli \metre} and height \SI{9.5}{\milli \metre} were added to each corner of the upper part of the slicer and subtracted from its lower part. A 3D surface mesh was finally generated from the binary image using the marching cubes algorithm and exported in \lstinline{.stl} format for printing. The brain slicer design is depicted in \Cref{fig:a1}.

Ten cutting guides were also designed to ease the cutting of the brain slices into blocks. Those consisted of $\SI{190.5}{\milli \metre} \times \SI{149.5}{\milli \metre} \times \SI{11}{\milli \metre}$ plates from which the slice volume with a thickness of \SI{5.5}{\milli \metre} was subtracted. Grooves were carved into the plates to guide the scalpel-cutting of $\SI{26}{\milli \metre} \times \SI{18}{\milli \metre} \times \SI{5.5}{\milli \metre}$ tissue blocks whose dimensions are compatible with the histological processing chain at our institution. All image processing and design steps were performed using an in-house software in C++ based on VTK \cite{schroeder_2010} and ITK \cite{yoo_2002}.

The brain slicer was printed using \SI{1.75}{\milli \metre} PLA thread on a HICTOP 3DD-17-ATL-FM printer (HIC Technology, China) with a nozzle size of \SI{0.4}{\milli \metre} (layer thickness \SI{0.2}{\milli \metre}, honeycomb filling \SI{5}{\percent}) for a total printing time of 96 hours. The printing of the ten cutting guides was parallelized on 5 Prusa i3 MK2 printers (Prusa Research, Czech Republic) with equivalent settings for a printing time of around 10 hours per guide.

\section{Deep learning-based nuclei counting}
\label{app:b}
For each of the 28 stained slides resampled to a pixel size of $\SI{1}{\micro \metre} \times \SI{1}{\micro \metre}$, 50 tiles of $\SI{100}{\pixel} \times \SI{100}{\pixel}$ were randomly chosen after exclusion of background tiles. Background tiles were defined as tiles containing less than \SI{10}{\percent} of tissue pixels, empirically chosen as pixels whose the lowest RGB value is above 220 for our calibrated scanner. The 1600-tile dataset was then weakly supervised using an in-house annotation program in Python allowing the user to locate each cell nucleus within a tile by a simple mouse click. For each annotated tile, a nuclei presence probability map was generated by the superposition of 2D Gaussian blobs centered on user-pointed nucleus locations with full width at half maximum (FWHM) of \SI{3}{\pixel} (i.e. standard deviation $\sigma \approx \SI{1.27}{\pixel}$), truncated to a radius of $4\,\sigma$. The supervised dataset was split into training and validation sets in proportion \SI{80}{\percent}--\SI{20}{\percent}.

\begin{figure}[ht!]
    \centering
    \includegraphics[width=\textwidth]{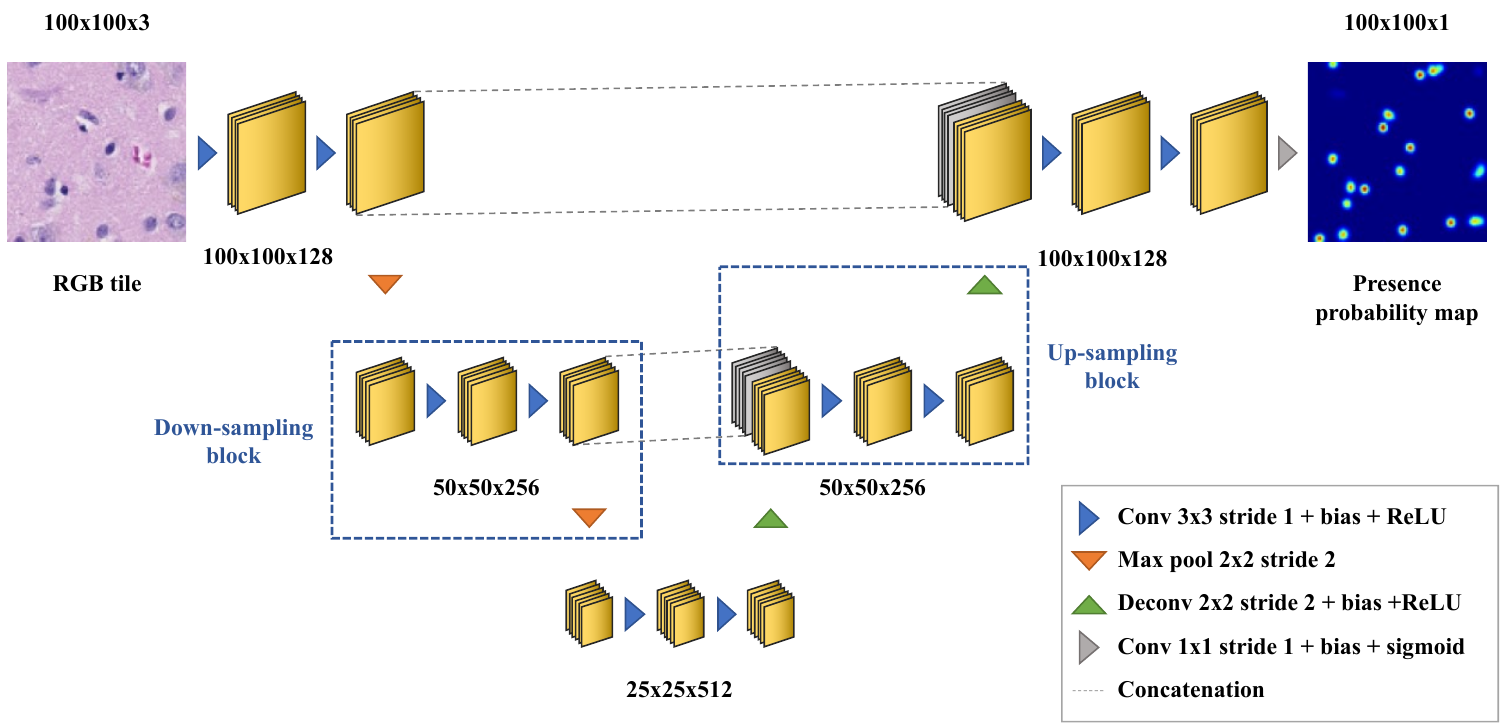}
    \caption{U-Net architecture with its feature map dimensions used for nuclei detection. The network takes $\SI{100}{\pixel} \times \SI{100}{\pixel}$ RGB tiles with pixel size $\SI{1}{\micro \metre} \times \SI{1}{\micro \metre}$ as input and predicts a nuclei presence probability map with the same spatial dimensions.}
    \label{fig:a2}
\end{figure}

A U-Net \cite{ronneberger_2015} deep convolutional neural network was implemented for cell nuclei detection, consisting of 2 down-sampling blocks, 3 up-sampling blocks and an output block. Each down-sampling block is made of two convolutional layers with kernel size $3 \times 3$ and stride 1 followed by a bias-adding layer and a rectified linear unit (ReLU) activation layer. A max pool layer with kernel size $2 \times 2$ and stride 2 is added at the end of the block to reduce the feature maps dimensions by a factor 2. The up-sampling blocks are identical to the down-sampling blocks except that the max pool layer is replaced by a deconvolution layer with kernel size $2 \times 2$ and stride 2 followed by a bias-adding layer and a ReLU activation layer to expand the feature maps dimensions by a factor 2. The output block has the same structure as the down-sampling blocks except that the max pool layer is replaced by a convolutional layer with kernel size $1 \times 1$ and stride 1 followed by a bias-adding layer and a sigmoid activation layer to merge the last 128 feature maps into a single presence probability map. The network architecture is depicted in \Cref{fig:a2} and was implemented in Python using the TensorFlow framework (version 1.13.1) \cite{abadi_2015}.

The network was trained on the weakly-supervised training set using the cross-entropy loss and the Adam optimizer \cite{kingma_2014} (learning rate: \num{e-4}, $\beta_1$: \num{0.9}, $\beta_2$: \num{0.999}, $\epsilon$: \num{e-6}). An evaluation was performed on the validation set at the end of each epoch and the training was stopped early after no improvement of the validation metric in 100 epochs. The network parameter values that gave the best validation metric value (\num{0.041}) were kept, which was reached at epoch 96. A similar weakly supervised deep-learning approach was used in \cite{xing_2019} for cell nuclei detection in histological slides, but Euclidian distance to user-pointed nuclei locations and the mean squared error loss were used instead of Gaussian nuclei presence probability maps and the cross-entropy loss.

After training, each resampled scanned slide was divided into adjacent tiles of $\SI{100}{\pixel} \times \SI{100}{\pixel}$ from which nuclei presence probability maps were predicted by the trained network. A nuclei count was finally derived for each predicted probability map by detecting its local maxima with a value greater or equal to \SI{0.1} and spaced by at least \SI{3}{\pixel}.

\section{\textit{In vivo} to \textit{ex vivo} registration}
\label{app:c}
Registration of the \textit{in vivo} T2-FLAIR image to the \textit{ex vivo} T1 image was performed using the Elastix software (version 4.801) \cite{klein_2010} as described in \Cref{subsec:2_7}. The Elastix parameter files for rigid and B-spline registration are respectively given in \Cref{lst:a1,lst:a2}

\begin{lstlisting}[backgroundcolor = \color{lightgray}, caption={Elastix parameter file for rigid \textit{in vivo} T1 to \textit{ex vivo} T1 registration.}, label={lst:a1}, language=bash, numbers=left]
// ElastixMain
(FixedImageDimension 3)
(MovingImageDimension 3)
(FixedInternalImagePixelType "double")
(MovingInternalImagePixelType "double")

// ElastixTemplate
(WriteTransformParametersEachIteration "false")
(WriteTransformParametersEachResolution "false")
(UseDirectionCosines "true")

// Registration
(Registration "MultiResolutionRegistration")
(NumberOfResolutions 1)

(ErodeMask "false")
(ErodeFixedMask "false")
(ErodeMovingMask "false")

// Pyramid
(FixedImagePyramid "FixedSmoothingImagePyramid")
(MovingImagePyramid "MovingSmoothingImagePyramid")
(ImagePyramidSchedule 1.0 1.0 1.0)
(WritePyramidImagesAfterEachResolution "false")

// Transform
(Transform "EulerTransform")
(AutomaticScalesEstimation "true")
(AutomaticTransformInitialization "true")
(AutomaticTransformInitializationMethod "GeometricalCenter")

// Metric
(Metric "AdvancedMattesMutualInformation")
(NumberOfHistogramBins 50)
(NumberOfFixedHistogramBins 50)
(NumberOfMovingHistogramBins 50)
(FixedKernelBSplineOrder 3)
(MovingKernelBSplineOrder 3)
(FixedLimitRangeRatio 0.01)
(MovingLimitRangeRatio 0.01)
(FiniteDifferenceDerivative "false")
(UseFastAndLowMemoryVersion "true")
(UseJacobianPreconditioning "false")
(ShowExactMetricValue "true")
(ExactMetricSampleGridSpacing 1 1 1)
(CheckNumberOfSamples "false")
(RequiredRatioOfValidSamples 0.25)

// Optimizer
(Optimizer "RegularStepGradientDescent")
(MaximumNumberOfIterations 500)
(MinimumStepLength 0.00001)
(MaximumStepLength 8.0)
(MinimumGradientMagnitude 0.00001)
(RelaxationFactor 0.8)
(NewSamplesEveryIteration "false")

// ImageSampler
(ImageSampler "Full")

// Interpolator
(Interpolator "LinearInterpolator")

// ResampleInterpolator
(ResampleInterpolator "FinalLinearInterpolator")

// Resampler
(Resampler "DefaultResampler")
(WriteResultImage "true")
(WriteResultImageAfterEachResolution "false") 
(WriteResultImageAfterEachIteration "false") 
(ResultImageFormat "mha")
(ResultImagePixelType "double")
(CompressResultImage "false")

// Misc
(UseMultiThreadingForMetrics "true")
(DefaultPixelValue 0.0)
\end{lstlisting}

\begin{lstlisting}[backgroundcolor = \color{lightgray}, caption={Elastix parameter file for B-Spline \textit{in vivo} T1 to \textit{ex vivo} T1 registration.}, label={lst:a2}, language=bash, numbers=left]
// ElastixMain
(FixedImageDimension 3)
(MovingImageDimension 3)
(FixedInternalImagePixelType "double")
(MovingInternalImagePixelType "double")

// ElastixTemplate
(WriteTransformParametersEachIteration "false")
(WriteTransformParametersEachResolution "false")
(UseDirectionCosines "true")

// Registration
(Registration "MultiResolutionRegistration")
(NumberOfResolutions 5)
(ErodeMask "false")
(ErodeFixedMask "false")
(ErodeMovingMask "false")

// Pyramid
(FixedImagePyramid "FixedSmoothingImagePyramid")
(MovingImagePyramid "MovingSmoothingImagePyramid")
(ImagePyramidSchedule 16.0 16.0 16.0 8.0 8.0 8.0 4.0 4.0 4.0 2.0 2.0 2.0 1.0 1.0 1.0)
(WritePyramidImagesAfterEachResolution "false")

// Transform
(Transform "BSplineTransform")
(BSplineTransformSplineOrder 3)
(FinalGridSpacingInPhysicalUnits 4.0 4.0 4.0)
(GridSpacingSchedule 16.0 16.0 16.0 8.0 8.0 8.0 4.0 4.0 4.0 2.0 2.0 2.0 1.0 1.0 1.0)
(PassiveEdgeWidth 0)
(UseCyclicTransform "false")
(HowToCombineTransforms "Compose")

// Metric
(Metric "AdvancedMattesMutualInformation")
(NumberOfHistogramBins 32)
(NumberOfFixedHistogramBins 32)
(NumberOfMovingHistogramBins 32)
(FixedKernelBSplineOrder 3)
(MovingKernelBSplineOrder 3)
(FixedLimitRangeRatio 0.001)
(MovingLimitRangeRatio 0.001)
(FiniteDifferenceDerivative "false")
(UseFastAndLowMemoryVersion "true")
(UseJacobianPreconditioning "false")
(ShowExactMetricValue "false")
(ExactMetricSampleGridSpacing 1)
(CheckNumberOfSamples "true")
(RequiredRatioOfValidSamples 0.25)

// Optimizer
(Optimizer "StandardGradientDescent")
(MaximumNumberOfIterations 12000) 
(MaximumNumberOfSamplingAttempts 10)
(SP_a 2500)
(SP_A 1200.0)
(SP_alpha 0.6) 
(NewSamplesEveryIteration "true")

// ImageSampler
(ImageSampler "RandomCoordinate")
(NumberOfSpatialSamples 2000)
(UseRandomSampleRegion "true")
(SampleRegionSize 100.0 100.0 100.0)
(FixedImageBSplineInterpolationOrder 1)

// Interpolator
(Interpolator "BSplineInterpolator")
(BSplineInterpolationOrder 1)

// ResampleInterpolator
(ResampleInterpolator "FinalBSplineInterpolator")
(FinalBSplineInterpolationOrder 3)

// Resampler
(Resampler "DefaultResampler")
(WriteResultImage "true")
(WriteResultImageAfterEachResolution "false") 
(WriteResultImageAfterEachIteration "false") 
(ResultImageFormat "mha")
(ResultImagePixelType "double")
(CompressResultImage "false")

// Misc
(UseMultiThreadingForMetrics "true")
(DefaultPixelValue 0.0)
\end{lstlisting}

\section{Pathology results}
\label{app:d}
The results of the pathological examination and numerical tile processing are summarized in \Cref{tab:a1}.

\begin{table}[H]
    \centering
    \caption[Results of the Pathological Examination and Numerical Tile Processing]{Results of the pathological examination and numerical tile processing. PPN: pseudo-palisading necrosis, GVP: glomeruloid vascular proliferation, susp.: suspected.}
    \resizebox{\textwidth}{!}{
    \begin{tabular}{rllllrrrrrr}
        \toprule
        & & & & & \multicolumn{3}{c}{\textbf{Cell density}} & \multicolumn{3}{c}{\textbf{Distance [\si{\milli \metre}]}} \\
        & & & & & \multicolumn{3}{c}{\textbf{[\SI{e+3}{\cell \per \square \milli \metre}]}} & & & \\
        & & & & & & \\
        \textbf{Block} & \textbf{PPN} & \textbf{Tumor cells} & \textbf{GVP} & \textbf{Edema} & \textbf{Min} & \textbf{Max} & \textbf{Mean} & \textbf{Min} & \textbf{Max} & \textbf{Mean} \\
        \midrule
        1  & No  & No               & No  & No  & 1.08 & 20.46 & 11.77 & 33.40 & 40.40 & 34.95 \\
        2  & No  & No               & No  & No  & 2.25 & 23.88 & 13.80 & 35.05 & 46.76 & 40.12 \\
        3  & No  & Infiltrative (susp.) & No  & Yes & 1.52 & 25.25 & 14.96 & 7.24  & 21.18 & 14.00 \\
        4  & No  & No               & No  & No  & 1.36 & 20.64 & 12.50 & 29.36 & 41.66 & 36.84 \\
        5  & No  & No               & No  & Yes & 1.19 & 24.85 & 15.40 & 11.74 & 30.27 & 21.57 \\
        6  & No  & Infiltrative (susp.) & No  & No  & 3.72 & 28.44 & 18.70 & 32.92 & 40.35 & 36.52 \\
        7  & No  & Infiltrative (susp.) & No  & Yes & 0.89 & 38.95 & 21.44 & 38.59 & 61.70 & 49.04 \\
        8  & Yes & Block            & Yes & No  & 3.37 & 37.29 & 23.65 & 0.50  & 5.17  & 2.71  \\
        9  & Yes & Infiltrative     & Yes & Yes & 1.02 & 44.37 & 31.71 & 0.50  & 4.72  & 1.91  \\
        10 & No  & Infiltrative (susp.) & No  & Yes & 5.52 & 37.15 & 25.53 & 0.50  & 3.71  & 1.06  \\
        11 & No  & No               & No  & No  & 1.73 & 25.68 & 15.21 & 19.85 & 38.74 & 30.74 \\
        12 & No  & Infiltrative     & Yes & Yes & 0.89 & 36.08 & 19.70 & 0.50  & 31.46 & 14.24 \\
        13 & Yes & Block            & Yes & No  & 1.40 & 52.46 & 21.69 & 0.50  & 22.85 & 9.72  \\
        14 & No  & No               & No  & No  & 1.10 & 19.25 & 11.78 & 17.73 & 32.20 & 25.51 \\
        15 & No  & No               & No  & Yes & 2.08 & 22.90 & 11.41 & 28.63 & 54.74 & 40.14 \\
        16 & No  & No               & No  & No  & 0.99 & 26.47 & 13.52 & 20.37 & 48.79 & 35.91 \\
        17 & No  & No               & No  & No  & 1.42 & 23.69 & 13.10 & 28.30 & 56.12 & 42.88 \\
        18 & No  & No               & Yes & Yes & 2.46 & 39.29 & 20.71 & 7.94  & 27.01 & 16.83 \\
        19 & Yes & Block            & Yes & Yes & 0.99 & 44.12 & 19.82 & 0.50  & 18.50 & 6.92  \\
        20 & Yes & Block            & Yes & No  & 6.25 & 61.05 & 28.40 & 0.50  & 7.37  & 2.06  \\
        21 & No  & Infiltrative     & Yes & Yes & 4.75 & 34.20 & 23.05 & 0.50  & 14.62 & 7.70  \\
        22 & No  & No               & No  & Yes & 0.86 & 30.37 & 10.79 & 2.99  & 36.87 & 21.39 \\
        23 & No  & No               & No  & No  & 0.94 & 25.61 & 13.23 & 21.14 & 49.61 & 34.48 \\
        24 & No  & No               & No  & No  & 1.37 & 26.72 & 13.72 & 57.27 & 90.93 & 71.08 \\
        25 & No  & Infiltrative (susp.) & No  & Yes & 0.87 & 39.07 & 21.03 & 1.71  & 21.41 & 10.87 \\
        26 & Yes & Block            & Yes & Yes & 6.21 & 54.24 & 22.79 & 0.50  & 8.81  & 2.96  \\
        27 & No  & No               & No  & Yes & 0.93 & 37.96 & 21.33 & 12.74 & 30.35 & 18.40 \\
        28 & No  & No               & No  & No  & 0.95 & 34.79 & 20.76 & 14.35 & 35.85 & 22.57 \\
        \bottomrule
    \end{tabular}}
    \label{tab:a1}
\end{table}

\end{document}